\documentclass[oneside,a4paper,12pt]{article}

\usepackage[utf8]{inputenc} 
\usepackage[T1]{fontenc}
\usepackage{microtype}
\usepackage{color, xcolor} 

\usepackage{abstract} 
\usepackage{authblk} 

\usepackage{titlesec} 
\usepackage[hidelinks]{hyperref} 
\usepackage{cite} 
\usepackage{appendix} 
\usepackage[titles]{tocloft}

\usepackage{multicol} 
\usepackage{fancyhdr} 
\usepackage{listings} 
\usepackage{graphicx, subfig} 
\usepackage{float} 
\usepackage{placeins}

\usepackage{enumerate} 
\usepackage{longtable} 
\usepackage{multirow} 
\usepackage{makecell} 
\usepackage{tabularx} 
\usepackage{array} 
\usepackage{booktabs}
\usepackage{bm}

\usepackage{amsfonts} 
\usepackage{amsmath} 
\usepackage{amssymb} 
\usepackage{gensymb} 
\usepackage{amsthm} 
\usepackage[ruled]{algorithm2e} 
\usepackage{tikz} 
\usepackage{tikz-cd}
\usetikzlibrary{decorations.markings}
\usetikzlibrary{arrows.meta}
\usetikzlibrary{positioning}
\usetikzlibrary{shapes.geometric}
\usetikzlibrary{fit}
\usetikzlibrary{backgrounds}

\usepackage{dsfont} 
\usepackage[normalem]{ulem} 
\usepackage{float} 

\usepackage[english]{babel}
\newtheorem{remark}{Remark}

\newtheoremstyle{breakhead}{\topsep}{\topsep}{\itshape}{}
  {\bfseries}{.}{\newline}{}
\theoremstyle{breakhead}

\theoremstyle{plain}

\usepackage[total={7in,9in}]{geometry}
\hypersetup{pdftitle={A NURBS-enhanced high-order finite-volume method for compressible laminar boundary layers on curved domains with hybrid meshes}}
\date{}

\begin{document}

\title{A NURBS-enhanced high-order finite-volume method for compressible
laminar boundary layers on curved domains with hybrid meshes}

\author[a]{Yisu Li}
\author[b,c]{Xucheng Meng}
\author[d]{Guanghui Hu\thanks{Corresponding author.}}
\affil[a]{Department of Mathematics, University of Macau, Macao SAR, China}
\affil[b]{Department of Mathematics, Faculty of Arts and Sciences,\protect\newline
Beijing Normal University, Zhuhai 519087, Guangdong, China}
\affil[c]{Guangdong Provincial Key Laboratory of Interdisciplinary Research and Application for Data Science,\protect\newline
Beijing Normal-Hong Kong Baptist University, Zhuhai 519087, China}
\affil[d]{State Key Laboratory of Internet of Things for Smart City and Department of Mathematics,
University of Macau, Macao SAR, China}

\maketitle

\begin{abstract}
  High-fidelity simulation of compressible laminar boundary layers on curved
  domains requires physical wall boundary treatment and geometric representation
  compatible with high-order reconstruction, since viscous shear stresses and
  conductive heat fluxes are concentrated in a thin near-wall region.
  Conventional high-order methods often suffer from loss of accuracy near curved
  walls or require complicated wall treatments.
  To address these issues, in this paper, we develop a NURBS-enhanced $k$-exact
  finite-volume method for the steady compressible Navier--Stokes equations on
  hybrid meshes.
  To incorporate physical wall conditions into the high-order
  discretization, we develop a wall boundary treatment that
  employs sequential reconstruction without requiring the
  reconstruction variables to coincide with those used to prescribe
  the boundary conditions.
  The NURBS-based boundary representation is used to obtain a high-order
  accurate description of curved boundaries, and the diffusive generalized Riemann problem (dGRP)
  viscous flux is employed to enhance numerical robustness.
  A variety of numerical examples are provided to demonstrate the high-order
  accuracy and effectiveness of the proposed method.
\end{abstract}


\newpage

\section{Introduction}

High-fidelity numerical simulation provides a foundation for understanding physical properties of
viscous flow and developing reliable predictive tools for aerodynamic
analysis and design~\cite{slotnick2014}. For wall-bounded compressible flows,
this predictive capability depends critically on resolving the thin boundary
layers in which near-wall momentum and heat transport determine surface
stresses and thermal loads. In the laminar regime, interactions between these
layers, pressure gradients, and surface geometry can produce separation and
reattachment, connecting localized viscous structures to the surrounding flow
and its aerodynamic response~\cite{mengaldo2015}. Resolving these interactions
poses a demanding numerical problem: strong wall-normal velocity and temperature
gradients must be represented on scales much smaller than those of the outer
flow, while curved boundaries and physical wall boundary conditions must be accurately represented.
Advancing the accurate and efficient computation of compressible laminar boundary layers therefore requires resolving both the near-wall solution and its geometric representation.

High-order computational fluid dynamics (CFD) methods reduce numerical dissipation and improve the
representation of smooth flow structures~\cite{wang2013high}.
Discontinuous Galerkin (DG) methods achieve high-order accuracy by representing
the solution with elementwise polynomial approximations and coupling neighboring
elements through numerical fluxes, thereby combining local conservation with the
flexibility of unstructured meshes~\cite{bassi1997,cockburn2001}.
Their extension to the compressible Navier--Stokes equations further requires an
appropriate treatment of viscous fluxes and interelement gradient coupling, as
demonstrated by Bassi and Rebay~\cite{bassi1997}.
Flux reconstruction and related correction formulations provide a unified
framework for several high-order numerical schemes, including discontinuous
Galerkin and spectral difference methods~\cite{huynh2007,wang2009}.
Within this framework, Navah and Nadarajah~\cite{navah2020verification}
used manufactured solutions to verify high-order accuracy on curved
wall-bounded domains.
To obtain high-order finite-difference methods for steady compressible Navier--Stokes
flows in curved domains,
Wan \emph{et al.}~\cite{wan2025curved}
combined a fifth-order hybrid weighted essentially non-oscillatory (WENO)
finite-difference scheme with WENO extrapolation and simplified inverse
Lax--Wendroff boundary treatment on Cartesian grids.
Reconstruction-based finite-volume methods obtain high-order solution and
gradient approximations from cell averages. These methods include $k$-exact reconstruction methods~\cite{barth1990,olliviergooch2002,michalak2009},
compact schemes on mixed grids with large aspect ratios~\cite{huang2022mixed},
and variational reconstruction methods~\cite{wang2017variational}.
These developments provide a solid foundation
for high-order compressible viscous-flow simulation.

In finite-volume methods, high-order solution
reconstruction alone does not ensure high-order accuracy
of a boundary-layer discretization on a curved domain. Three interconnected
numerical requirements must be addressed together: (1) accurate curved-wall
geometry with strongly anisotropic near-wall resolution, (2) viscous fluxes
that couple reconstructed states and gradients across cell interfaces,
and (3) physical wall boundary treatment that connects physical wall conditions
to the reconstruction and numerical boundary fluxes. Each affects the
near-wall stresses and heat fluxes, motivating their treatment within a common formulation.

Geometric fidelity and directional mesh resolution address complementary
parts of the boundary-layer problem. Errors in wall location and normal
direction affect boundary flux integration, while inaccurate curved-cell
moments enter the reconstruction. An accurate description of curved boundaries is critical to
designing high-order methods for the simulation of compressible flows on complex domains; see~\cite{wang2013high}
and references therein for details. Several methods have been proposed in the literature to reduce geometric errors.
For instance,
isogeometric analysis uses non-uniform rational B-splines (NURBS) for both
geometric representation and solution approximation~\cite{hughes2005},
thereby retaining the geometry of computer-aided design. NURBS-enhanced
finite element methods were subsequently developed for elliptic
problems~\cite{sevilla2008} and the compressible Euler
equations~\cite{Sevilla2008Euler}. Meng and Hu introduced a NURBS-enhanced
finite-volume method for the steady Euler equations~\cite{menghu2018} and improved its computational efficiency through
goal-oriented $h$-adaptivity~\cite{menghu2022adaptive}. On the resolution
side, resolving the thin near-wall viscous boundary layers requires a much finer
mesh size in the wall-normal direction than in the wall-parallel
directions or in the outer flow. Curvilinear reconstruction on anisotropic
meshes~\cite{jalali2017rans} and compact reconstruction on mixed grids with
large aspect ratios~\cite{huang2022mixed} address the associated accuracy
and conditioning requirements. These two aspects motivate the development of a 
high-order finite-volume method on curved domains, with stretched 
quadrilateral cells in the near-wall region and unstructured triangular cells in the outer flow region.

\begingroup
In the compressible Navier--Stokes equations, viscous stresses and heat fluxes
depend on both the local flow state and its spatial gradients. Independent
cellwise reconstructions generally have discontinuous interface traces.
Simply averaging the one-sided gradients does not explicitly account for
the state jump and can leave insufficient intercell coupling, including
possible odd--even decoupling~\cite{jalali2014diffusive,sejekan2016diffusive}.
Jalali \emph{et al.}~\cite{jalali2014diffusive} compared cell-centered
diffusive-flux discretizations on unstructured meshes and found that
adding a solution-jump term to the averaged gradients improved flux
accuracy and damping.
Sejekan and Ollivier-Gooch~\cite{sejekan2016diffusive} proposed a modified
least-squares reconstruction and interface-gradient evaluation to improve
the accuracy of diffusive fluxes and their integrals.
More recently, Setzwein \emph{et al.}~\cite{setzwein2024viscous}
developed an orthogonal--nonorthogonal viscous-flux decomposition for
vertex-centered $k$-exact finite-volume schemes, treating both interior
faces and Dirichlet boundaries.
The diffusive generalized Riemann problem (dGRP) formulation of Gassner
\emph{et al.}~\cite{gassner2007}, in the reconstruction-based form of
Wang \emph{et al.}~\cite{wang2017variational}, corrects the averaged
one-sided gradient by a face-normal term proportional to the state jump.
We adopt this established viscous-flux formulation to strengthen
intercell coupling between the conservative reconstructions on the
hybrid mesh and enhance numerical robustness.
\par
\endgroup

High-order wall boundary treatment has an established history in
reconstruction-based finite-volume methods. Boundary constraints are
added to the least-squares reconstruction for the advection--diffusion equation~\cite{olliviergooch2002}
and compressible Navier--Stokes flow~\cite{michalak2009}; the latter
includes physical wall data as reconstruction goals. Hoshyari~\cite{hoshyari2017}
describes reconstruction constraints and boundary fluxes for adiabatic walls
and symmetry boundaries. In the simulation of cylindrical Couette flow, Michalak
and Ollivier-Gooch~\cite{michalak2009} noted that the inner-cylinder
isothermal condition was difficult to enforce with their chosen
reconstruction variables. They supplemented the prescribed wall velocity
with an isobaric condition to obtain an equivalent result for that
configuration. However, this substitution is specific to the 
Couette-flow configuration and is not generally applicable to isothermal wall 
boundaries. Subsequent multiple-correction $k$-exact schemes
incorporate wall data, curved geometry, and enriched
stencils~\cite{menasria2018,menasria2019}. Costa \emph{et al.}~\cite{costa2019}
extended reconstruction for off-site data (ROD) to general boundary
conditions on smooth curved boundaries for the convection--diffusion equation.
In the quadratic-reconstruction Couette calculation of Menasria
\emph{et al.}~\cite{menasria2018}, the expected viscous convergence order
was not attained, and the calculation of boundary diffusive fluxes was identified as an area for further
work. These developments show that reconstruction constraints and the
corresponding wall fluxes must be considered together. The present wall
treatment incorporates physical wall conditions into the $k$-exact reconstruction
of conservative variables and uses reconstructed wall states and one-sided viscous
gradients to evaluate numerical fluxes for no-slip isothermal and adiabatic walls.

This work develops a NURBS-enhanced
high-order $k$-exact finite-volume method for the steady compressible
Navier--Stokes equations on curved domains with triangular--quadrilateral hybrid meshes.
The formulation combines NURBS boundary integration and curved-cell
moments, stretched quadrilateral layers with unstructured triangles,
and reconstruction.
Within this framework, the proposed wall boundary treatment connects prescribed
velocity and thermal data to the reconstructed solution and numerical wall fluxes. The main contributions are:

\begin{itemize}

\item \textbf{High-order wall boundary treatment for $k$-exact finite-volume methods.}
We develop a high-order wall boundary treatment that incorporates physical wall conditions
into the $k$-exact reconstruction of conservative variables.
In finite-volume methods, the degrees of freedom are conservative cell averages, whereas physical wall conditions are typically prescribed in primitive variables.
The nonlinear relationship between these variables complicates their coupling within the reconstruction system.
To address this difficulty, we introduce a sequential reconstruction
strategy in which density is reconstructed first and its wall values are then used
to construct momentum targets and, for isothermal walls, total-energy targets at boundary quadrature points.
These targets are incorporated as additional goals in the weighted least-squares reconstruction.
The resulting treatment allows physical wall conditions to be incorporated without requiring the reconstruction variables to coincide with those used to prescribe them.

\item \textbf{An efficient high-order finite-volume framework for compressible
boundary layers on curved domains with hybrid meshes.}
We propose a high-order NURBS-enhanced finite-volume method on curved domains with triangular--quadrilateral
hybrid meshes for solving the compressible Navier--Stokes equations. Stretched quadrilateral layers provide directional near-wall
resolution, while the unstructured meshes preserve flexibility in the outer flow. The NURBS-enhanced curved boundary description is used to provide accurate geometric information for $k$-exact reconstruction on curved cells and flux integration along the curved wall boundary. To efficiently solve the resulting nonlinear system,
we employ a pseudo-transient matrix-free Newton--Krylov method with thin-layer preconditioning,
avoiding the assembly of the full high-order Jacobian and enabling efficient steady-state computation. 

\end{itemize}


Section~2 states the governing equations and finite-volume framework.
Section~3 develops reconstruction, interior fluxes, and physical boundary
treatment. Section~4 describes hybrid meshes and fitted geometry, and
Section~5 presents the steady solver. Section~6 reports verification
and boundary-layer applications, followed by conclusions in Section~7.

\section{Governing Equations and Finite-Volume Discretization}

\subsection{Governing equations}

We consider the two-dimensional steady compressible Navier--Stokes
equations for laminar flow of an ideal gas. In conservative form, they read
\begin{equation}
\nabla \cdot \bigl[\mathcal F^c(U)
-\mathcal F^v(U,\nabla U)\bigr]=\bm 0,\quad \mbox{in} \,\, \Omega,
\label{eq:nseq}
\end{equation}
where $U=(\rho,\rho u,\rho v,E)^T$ is the vector of
conservative variables. Here $\rho$, $\vec v=(u,v)^T$, $P$, and $E$
denote the density, velocity, pressure, and total energy per unit volume, respectively, and
$\nabla U=[\partial_x U,\partial_y U]\in\mathbb R^{4\times2}$ is the gradient of $U$.
The convective flux tensor is
\begin{equation}
\mathcal F^c(U)
=
\begin{bmatrix}
\rho u          & \rho v          \\[2pt]
\rho u^2 + P    & \rho uv         \\[2pt]
\rho uv         & \rho v^2 + P    \\[2pt]
u(E + P)        & v(E + P)
\end{bmatrix},
\label{eq:convective_flux}
\end{equation}
and the viscous flux tensor is
\begin{equation}
\mathcal F^v(U,\nabla U)
=
\begin{bmatrix}
0 & 0 \\[2pt]
\tau_{xx} & \tau_{xy} \\[2pt]
\tau_{yx} & \tau_{yy} \\[2pt]
u\tau_{xx} + v\tau_{xy} - q_x
&
u\tau_{xy} + v\tau_{yy} - q_y
\end{bmatrix},
\label{eq:viscous_flux}
\end{equation}
where $\tau$ is the viscous stress tensor with Cartesian components
$\tau_{xx}$, $\tau_{xy}$, $\tau_{yx}$, and $\tau_{yy}$,
and $\vec q=(q_x,q_y)^T$ is the heat flux. The ideal-gas equation of
state closes the system:
\begin{equation}
P=(\gamma-1)\left(E-\tfrac12\rho\|\vec v\|_2^2\right)
  = \rho RT,
\label{eq:eos}
\end{equation}
where $T$ is temperature, $R$ is the specific gas constant, and
$\gamma=1.4$ is the ratio of specific heats. The specific heats are
$c_v=R/(\gamma-1)$ and $c_p=\gamma c_v$.
For a Newtonian fluid with Stokes' hypothesis, the stress tensor and
Fourier heat flux are
\begin{equation}
\tau
= \mu \Bigl(
\nabla\vec v + (\nabla\vec v)^T
- \tfrac{2}{3}(\nabla\cdot\vec v)\,I_2
\Bigr),
\label{eq:stress}
\end{equation}
\begin{equation}
\vec q=-\kappa\nabla T,
\label{eq:heat_flux}
\end{equation}
in which $\mu$ is dynamic viscosity, $I_2$ is the two-dimensional
identity tensor, and $\kappa=\mu c_p/\mathrm{Pr}$ is thermal conductivity,
where $\mathrm{Pr}$ is the Prandtl number. For Couette flow, we use constant
viscosity and $\mathrm{Pr}=0.71$. In all other cases, viscosity follows
Sutherland's law and $\mathrm{Pr}=0.72$.

\subsection{Cell-centered finite-volume discretization}

Let the physical domain $\Omega\subset\mathbb{R}^2$ be partitioned into a hybrid mesh consisting of $N$
non-overlapping triangular and quadrilateral control volumes, that is,
$\overline\Omega=\bigcup_{i=1}^N\overline{\Omega_i}$.
Let $A_i=|\Omega_i|$ and let $\overline U_i\in\mathbb R^4$ denote the cell average of $U$ on $\Omega_i$, with
$\overline U=(\overline U_1^T,\ldots,
\overline U_N^T)^T\in\mathbb R^{4N}$.
Integrating Eq.~\eqref{eq:nseq}
over $\Omega_i$ and applying the divergence theorem gives
\begin{equation}
\sum_{e \subset \partial\Omega_i}
\int_e \bigl[\mathcal F^c(U)
-\mathcal F^v(U,\nabla U)\bigr]
\cdot\hat n\,\mathrm{d}s
=0,
\label{eq:semidisc}
\end{equation}
where $e$ is an edge (face) of the control volume $\Omega_i$, and $\hat n$ is the corresponding outward unit normal.

Replacing the physical fluxes by numerical fluxes and applying Gaussian quadrature to each edge yields the discrete residual:
\begin{equation}
R_i(\overline U)
=\frac{1}{A_i}
\sum_{e\subset\partial\Omega_i}
\sum_{q=1}^{N_e} w_q
\left(
F_q^c
-F_q^v
\right),
\label{eq:discrete_residual}
\end{equation}
where $N_e$ is the number of numerical quadrature points on edge $e$, $w_q$ are the quadrature weights, and
\begin{equation}
\begin{aligned}
F_q^c
&=F^c\left(U_L,U_R,\hat n_q\right),\\
F_q^v
&=F^v\left(U_L,U_R,\nabla U_L,\nabla U_R,\hat n_q\right),
\end{aligned}
\label{eq:generic_face_fluxes}
\end{equation}
where $U_L, U_R$ are left and right states, obtained by evaluating the reconstructed polynomials from the left and right 
cells at the quadrature point $\vec x_q$ on edge $e$, and $\hat n_q$ is the unit outward normal to $e$ at $\vec x_q$, pointing from the left cell to the right cell.
For an edge on a physical boundary, the reconstructed interior state and its gradient, together with the prescribed boundary conditions, are used to compute the numerical fluxes.
Collecting the residuals gives the
nonlinear system
\begin{equation}
R(\overline U)
=(R_1^T,\ldots,R_N^T)^T
=0.
\label{eq:ode}
\end{equation}

The nonlinear system~\eqref{eq:ode} can be solved using various iterative methods, including the full approximation scheme (FAS) multigrid
method~\cite{koren1990}, explicit high-order time
integration~\cite{wan2025curved}, and Newton--Krylov
methods~\cite{nejat2008}. In the present work, we employ a pseudo-transient
matrix-free Newton--Krylov method with thin-layer preconditioning, as presented
in Section~5.

\section{Reconstruction, Numerical Fluxes, and Boundary Conditions}

\subsection[k-exact reconstruction]{$k$-exact reconstruction}
\label{sec:kexact}

In this paper, we consider $k$-exact least-squares reconstruction with $1\le k \le 3$. Following Michalak and Ollivier-Gooch~\cite{michalak2009}, a degree-$k$ polynomial is reconstructed from neighboring control-volume averages,
with nominal order $k+1$ for smooth solutions. For a scalar function $f(\vec x)$, let
$\overline f_i$ be its cell average over $\Omega_i$, evaluated by numerical quadrature for both polygonal and curved control volumes:
\begin{equation}
\overline f_i=
\frac{\sum_q w_{iq}f(\vec x_{iq})}{\sum_q w_{iq}},
\label{eq:average}
\end{equation}
where $\vec x_{iq}$ and $w_{iq}$ are the numerical quadrature points and weights in $\Omega_i$, respectively.

Let $\vec x_i=(x_i,y_i)^T$ be the reference point of control volume $\Omega_i$, with
local coordinate scales $h_{i,x}$ and $h_{i,y}$. For a multi-index
$\alpha=(\alpha_1,\alpha_2)$, where $\alpha_1,\alpha_2$ are nonnegative integers, we define
\begin{equation}
\psi_{i,\alpha}(\vec x)=
\frac{(x-x_i)^{\alpha_1}(y-y_i)^{\alpha_2}}
{h_{i,x}^{\alpha_1}h_{i,y}^{\alpha_2}}, \quad |\alpha| = \alpha_1 + \alpha_2  \le k.
\label{eq:normalized_monomials}
\end{equation}
We define the basis functions as
\begin{equation}
\phi_{i,\alpha}(\vec x)=\psi_{i,\alpha}(\vec x)
-\overline{\psi_{i,\alpha}}_i,
\qquad 1\leq|\alpha|\leq k,
\label{eq:monomials}
\end{equation}
where $\overline{\psi_{i,\alpha}}_i$ is the cell average of 
$\psi_{i,\alpha}$ on $\Omega_i$. 
We define the reconstructed polynomial of degree $k$ on $\Omega_i$ by
\begin{equation}
f_i(\vec x)=\overline f_i+
\sum_{1\leq|\alpha|\leq k}a_{i,\alpha}
\phi_{i,\alpha}(\vec x).
\label{eq:recon}
\end{equation}
The reconstructed conservative-variable field $U_i(\vec x)$ is obtained by applying Eq.~\eqref{eq:recon} componentwise to $U=(\rho,\rho u,\rho v,E)^T$.
The unknown reconstruction coefficients $a_{i,\alpha}\in\mathbb R$
can be determined as follows. 
First, construct a reconstruction stencil $\mathcal T_i$ for the control volume $\Omega_i$. The stencil $\mathcal T_i$ is formed from successive layers of face neighbors, with additional inward cells for cubic wall reconstruction. Then the coefficients $a_{i,\alpha}$ are chosen to match the neighboring control-volume averages,
\begin{equation}
w_{ij}\overline{f_i}_j = w_{ij} \overline f_j,
\qquad \forall \Omega_j\in\mathcal T_i, \,\, j \ne i,
\label{eq:kexact_constraint}
\end{equation}
in a weighted least-squares sense, where $\overline{f_i}_j$ is the cell average of $f_i(\vec x)$ over $\Omega_j$, and $w_{ij} = 1/\| \vec{x}_j - \vec{x}_i \|_2$ is the weight. 

Let $W(U)=(\rho,u,v,P)^T$ denote the primitive variables, with pressure
as the thermodynamic variable.
Unlike Michalak and Ollivier-Gooch~\cite{michalak2009}, who reconstruct
primitive variables, the present method performs $k$-exact reconstruction
directly on conservative variables. The implications of this choice for
physical boundary conditions are discussed in Section~\ref{sec:wall_treatment}.

\subsection{Convective and viscous flux evaluation}
\label{sec:viscous_flux}

We use Roe or HLLC numerical fluxes~\cite{roe1981,toro1994} to compute the convective flux, as listed
in Table~\ref{tab:case_configuration}. At each interior-face Gauss point, the dGRP interface gradient~\cite{gassner2007,wang2017variational} is given by
\begin{equation}
(\nabla U)^*=\tfrac12(\nabla U_L+\nabla U_R)
+\sigma_e(U_R-U_L)\otimes\hat n_q,
\label{eq:dgrp}
\end{equation}
where $\hat n_q$ points from $\Omega_L$ to $\Omega_R$, and
\begin{equation}
\sigma_e=\frac{1}{2\widetilde h_e},\qquad
\widetilde h_e=\frac{\min(A_L,A_R)}{|e|}.
\label{eq:dgrp_penalty}
\end{equation}
The chain rule converts $(\nabla U)^*$ to the velocity gradient
$(\nabla\vec v)^*$ and temperature gradient $(\nabla T)^*$ at
$W^*=(W_L+W_R)/2$. Let $\vec v^*=(\vec v_L+\vec v_R)/2$,
$\mu_q=[\mu(T_L)+\mu(T_R)]/2$, and $\kappa_q=\mu_q c_p/\mathrm{Pr}$.
Using Eqs.~\eqref{eq:stress}, \eqref{eq:heat_flux}, and
\eqref{eq:viscous_flux}, the interface stress, heat flux, and numerical
normal viscous flux are
\begin{equation}
\begin{aligned}
\tau_q&=\mu_q\Bigl(
(\nabla\vec v)^*+\bigl((\nabla\vec v)^*\bigr)^T\\
&\qquad\quad-\tfrac{2}{3}(\nabla\cdot\vec v)^*\,I_2
\Bigr),\\
\vec q&=-\kappa_q(\nabla T)^*,\\
F_q^v&=\mathcal F_q^v\cdot\hat n_q\\
&=\begin{bmatrix}
0\\
\tau_q\hat n_q\\
(\vec v^*)^T\tau_q\hat n_q-\vec q\cdot\hat n_q
\end{bmatrix}.
\end{aligned}
\label{eq:interior_viscous_flux}
\end{equation}
\begingroup
For comparison, the conventional viscous-flux calculation uses the
averaged gradient
\begin{equation}
(\nabla U)^{\mathrm{av}}=\tfrac12(\nabla U_L+\nabla U_R)
\label{eq:averaged_gradient}
\end{equation}
at each interior-face Gauss point. The chain rule at $W^*$ converts this
averaged conservative gradient to the velocity and temperature gradients
used in Eq.~\eqref{eq:interior_viscous_flux}.
The normal state-jump term in Eq.~\eqref{eq:dgrp} enters the momentum flux
through $\tau_q$ and the energy flux through both $\tau_q$ and $\vec q$,
providing explicit coupling between the reconstructed interface states
and gradients~\cite{wang2017variational}.
\par
\endgroup

\subsection{Boundary conditions}
\label{sec:wall_treatment}
\label{sec:physical_boundaries}

In our previous work on the Euler equations~\cite{menghu2018,menghu2022adaptive},
solid-wall and inflow/outflow boundary conditions were imposed directly through
numerical boundary fluxes, without being incorporated into the reconstruction
system, and the expected order of accuracy was achieved. In the present work on
the compressible Navier--Stokes equations, however, we find that imposing
boundary conditions solely through numerical boundary fluxes does not yield
correct solutions for viscous wall problems. Earlier high-order finite-volume
studies have also demonstrated the importance of incorporating physical boundary
conditions into reconstruction~\cite{olliviergooch2002,michalak2009}. Following
these studies, we therefore incorporate the corresponding wall boundary
conditions into the weighted least-squares reconstruction as additional goals at
boundary Gauss quadrature points.

The difficulty arises from the choice of reconstruction variables.
Physical wall boundary conditions are generally prescribed in terms of primitive
variables, such as velocity and temperature, whereas the present $k$-exact
reconstruction is performed directly on conservative variables. In finite-volume
methods, the degrees of freedom are the cell averages of conservative variables.
Since the transformation $W(U)=(\rho,u,v,P)^T$ between conservative and primitive
variables is nonlinear, it does not generally commute with control-volume averaging:
\begin{equation}
W(\overline{U}_i)\ne
\overline{W(U)}_i.
\label{eq:conversion_noncommutation}
\end{equation}
Consequently, primitive-variable boundary conditions cannot be imposed as
independent Dirichlet conditions on the reconstructed conservative variables
without accounting for the relations between them.

Michalak and Ollivier-Gooch~\cite{michalak2009} incorporated boundary conditions
into $k$-exact reconstruction by introducing additional reconstruction goals at
boundary Gauss quadrature points. Their method reconstructs the primitive
variables $(\rho,u,v,P)$, allowing the prescribed no-slip
velocities to be used directly as reconstruction goals. Although this choice avoids the need to convert prescribed
wall velocities into conservative momentum constraints, their formulation still 
does not directly accommodate isothermal wall conditions. In the simulation of cylindrical Couette
flow, they therefore resorted to an isobaric condition in place of
the prescribed wall temperature, relying on an equivalence specific to that
problem.

In this work, we extend the least-squares treatment of boundary
conditions in~\cite{michalak2009} to the direct reconstruction of conservative
variables. Using the reconstructed wall density, we construct momentum
targets from prescribed wall velocities and, for isothermal walls, total-energy
targets from prescribed temperatures.
These targets are incorporated as additional goals in the weighted
least-squares reconstruction.
This treatment allows physical wall conditions to be incorporated
without requiring the reconstruction variables to coincide with those used to prescribe them.

For a fixed control volume $\Omega_i$ whose reconstruction stencil reaches
the viscous wall boundary, we augment
the interior reconstruction system~\eqref{eq:kexact_constraint} with goals at the
boundary Gauss quadrature points. In the pointwise formulas below, spatial arguments
are suppressed: $f_i$ and $\phi_{i,\alpha}$ denote their values
at $\vec x_q$, and boundary targets are evaluated at the same point.
For each conservative component $f$ constrained by the wall conditions, we solve
the augmented system
\begin{equation}
\left\{
\begin{aligned}
w_{ij}\overline{f_i}_j&=w_{ij}\overline f_j,
&&\Omega_j\in\mathcal T_i,\ j\ne i,\\
w_{iq}^{\mathrm B}f_i&=w_{iq}^{\mathrm B}f_b,
&&\vec x_q\in\mathcal Q_i^{\mathrm B},
\end{aligned}
\right.
\label{eq:wall_augmented}
\end{equation}
in a weighted least-squares sense. Here $\mathcal Q_i^{\mathrm B}$ is the set
of wall Gauss points selected during stencil construction for $\Omega_i$,
including those on neighboring boundary cells; $f_b$ is the boundary target
constructed from the prescribed wall data and reconstructed density,
and $w_{iq}^{\mathrm B}=\|\vec x_q-\vec x_i\|_2^{-1}$ is the boundary row
weight. The scalar wall residual is $\mathcal B_q^{\mathrm w}=f_i-f_b$. The
boundary conditions are thus incorporated as additional reconstruction goals in
the weighted least-squares problem. Because the basis functions in
Eq.~\eqref{eq:monomials} have zero cell average, modifying only the nonconstant
coefficients in Eq.~\eqref{eq:recon} leaves the prescribed cell average unchanged.

\paragraph{Viscous walls} The no-slip condition prescribes the wall velocity
$\vec v_w$, which determines the conservative momentum
$m=\rho\vec v$ through the local wall density. For adiabatic
walls, only the momentum components are constrained in the reconstruction,
and the zero normal heat-flux condition is imposed through the viscous
boundary flux; total energy is reconstructed using the interior problem~\eqref{eq:kexact_constraint}.
For an isothermal wall, the prescribed temperature $T_w$ additionally
determines the conservative total energy $E$ through the equation of state,
providing an additional energy reconstruction target. 

For viscous walls, we employ sequential reconstruction. First, we reconstruct
density using the interior least-squares problem~\eqref{eq:kexact_constraint}, without
additional wall constraints. Once the density has been reconstructed, its
trace at each wall Gauss point $\vec x_q$, denoted by $\rho_q^{\mathrm w}$,
is obtained from the adjacent boundary cell. The prescribed wall velocity
$\vec v_w$ and temperature $T_w$ then define the following conservative boundary
targets for momentum and total energy:
\begin{equation}
\begin{aligned}
m_b&=\rho_q^{\mathrm w}\vec v_w,\\
E_b
&=\rho_q^{\mathrm w}
\left(c_vT_w+\frac12\|\vec v_w\|_2^2\right)
\qquad\text{(isothermal)}.
\end{aligned}
\label{eq:wall_goals}
\end{equation}

The momentum and, for isothermal walls, total-energy components are then
reconstructed with these density-dependent targets held fixed. For each constrained
component, the boundary residual is
\begin{equation}
\begin{aligned}
\mathcal B_q^{\mathrm w}
&=f_i-f_b\\
&=\overline f_i+
\sum_{1\leq|\alpha|\leq k}
a_{i,\alpha}\phi_{i,\alpha}-f_b,
\end{aligned}
\label{eq:wall_boundary_residual}
\end{equation}
where $f_b$ is the corresponding component of $m_b$ or $E_b$ in
Eq.~\eqref{eq:wall_goals}.
For these components, the boundary rows in
Eq.~\eqref{eq:wall_augmented} can be written as
\begin{equation}
w_{iq}^{\mathrm B}
\sum_{1\leq|\alpha|\leq k}a_{i,\alpha}\phi_{i,\alpha}
=w_{iq}^{\mathrm B}(f_b-\overline f_i).
\label{eq:wall_boundary_matrix}
\end{equation}
The fixed targets make these residuals linear in the coefficients, so the
constrained components are reconstructed separately through
Eq.~\eqref{eq:wall_augmented}.

\begingroup
At each wall Gauss point, the chain rule converts the reconstructed
interior state $U_L$ and its reconstructed gradient $\nabla U_L$ to the one-sided
velocity and temperature gradients $\nabla\vec v_L$ and $\nabla T_L$.
The exterior state $U_R$ is constructed with $\rho_R=\rho_L$ and
$\vec v_R=2\vec v_w-\vec v_L$. For an isothermal wall,
$T_R=2T_w-T_L$; for an adiabatic wall, $E_R=E_L$.
The transport coefficients $\mu_q$ and $\kappa_q$ are evaluated from these
states as in Section~\ref{sec:viscous_flux}.
The velocity gradient in the wall viscous flux is $\nabla\vec v_L$, and
the temperature gradient is
\begin{equation}
(\nabla T)^{\mathrm w}=
\begin{cases}
\nabla T_L, & \text{isothermal},\\
\nabla T_L-(\nabla T_L\cdot\hat n_q)\hat n_q,
& \text{adiabatic}.
\end{cases}
\label{eq:wall_temperature_gradient}
\end{equation}
Using Eq.~\eqref{eq:stress} with $\mu_q$ and $\nabla\vec v_L$ gives the
wall stress $\tau_q$, while $\vec q=-\kappa_q(\nabla T)^{\mathrm w}$.
The numerical normal viscous flux at the wall quadrature point is
\begin{equation}
F_q^v=
\begin{bmatrix}
0\\
\tau_q\hat n_q\\
\vec v_w^T\tau_q\hat n_q-\vec q\cdot\hat n_q
\end{bmatrix}.
\label{eq:wall_viscous_flux}
\end{equation}
For an adiabatic wall, Eq.~\eqref{eq:wall_temperature_gradient} gives
$\vec q\cdot\hat n_q=0$.
\par
\endgroup

\paragraph{Open boundaries} At subsonic inflow/outflow boundaries,
incoming inviscid characteristic information must be combined with viscous
effects. The Euler equations have three incoming characteristic modes at
inflow and one at outflow~\cite{thompson1987,giles1990,poinsot1992}, whereas
the two-dimensional Navier--Stokes equations require four inflow and three
outflow boundary conditions. Following the boundary flux formulation of
Huang \emph{et al.}~\cite{huang2018wellposed}, we combine the boundary flux correction
with the normal viscous flux in Robin conditions and augment the reconstruction
systems with corresponding goals at boundary Gauss quadrature points.

Let $W_b=(\rho_b,u_b,v_b,P_b)^T$ be the
prescribed freestream state and $U_b$ its conservative counterpart.
At $\vec x_q$, write $U=U_i(\vec x_q)$,
$\nabla U=\nabla U_i(\vec x_q)$, and $\hat n=\hat n_q$.
Define the unit tangent $\hat t=(-\hat n_y,\hat n_x)^T$, the velocity
components $v_n=\vec v\cdot\hat n$ and $v_t=\vec v\cdot\hat t$, and the
sound speed $a=\sqrt{\gamma P/\rho}$.
For the prescribed velocity $\vec v_b=(u_b,v_b)^T$ and sound speed $a_b$,
let $\Delta\vec v=\vec v-\vec v_b$.
With the prescribed state held fixed, the boundary flux correction is defined by
\[
b(U)=\!
\begin{cases}
\bigl[\mathcal F^c(U)-\mathcal F^c(U_b)\bigr]\cdot\hat n,&v_n\leq-a,\\
b_{\rm in}(U),&-a<v_n\leq0,\\
(1-v_n^2/a^2)(P-P_b)(0,\hat n^T,v_n)^T,&0<v_n<a,\\
0,&v_n\geq a
\end{cases}.
\]
For subsonic inflow, define
\[
\begin{aligned}
\Delta s&=\ln\frac{P}{P_b}-\gamma\ln\frac{\rho}{\rho_b},\\
\chi&=\Delta\vec v\cdot\hat n-\frac{2(a-a_b)}{\gamma-1}
+\frac{a\Delta s}{\gamma(\gamma-1)},\\
b_{\rm in}&=v_n\Delta s\,r_s
+v_n(\Delta\vec v\cdot\hat t)r_t
+(v_n-a)\chi r_-,
\end{aligned}
\]
where
\[
\begin{aligned}
r_s&=-\frac{\rho}{\gamma}
\bigl(1,\vec v^T,\tfrac12\|\vec v\|_2^2\bigr)^T,\\
r_t&=\rho(0,\hat t^T,v_t)^T,\\
r_-&=\frac{\rho}{2}
\begin{pmatrix}
-1/a\\
\hat n-\vec v/a\\
v_n-a/(\gamma-1)-\|\vec v\|_2^2/(2a)
\end{pmatrix}.
\end{aligned}
\]
The three terms in $b_{\rm in}$ correspond to the entropy, tangential-velocity,
and incoming acoustic characteristic modes. The subsonic outflow branch uses
the pressure correction of Huang \emph{et al.}~\cite{huang2018wellposed}.
The additional reconstruction residual for the Robin condition
$\mathcal B_q=0$ is the four-component vector
\begin{equation}
\mathcal B_q=b(U)-\mathcal F^v(U,\nabla U)\cdot\hat n.
\label{eq:open_boundary_residual}
\end{equation}
At the subsonic inflow boundary, all four components of $\mathcal B_q$ are
included as additional reconstruction goals. At the subsonic outflow boundary,
the two momentum components and the total-energy component are included. The
mass residual at the subsonic outflow boundary vanishes because both $b(U)$ and
the Navier--Stokes viscous flux have zero mass components.

The selected components of Eq.~\eqref{eq:open_boundary_residual} are minimized
together with the interior reconstruction residuals in a vector-valued, coupled
extension of the weighted least-squares system~\eqref{eq:wall_augmented}.
Their dependence on the conservative
variables and their spatial derivatives couples the reconstruction coefficients
of all four variables, which are therefore determined simultaneously. The same
polynomial representation~\eqref{eq:recon} is used.

On these open faces, the total numerical boundary flux in the finite-volume
residual is evaluated using the same boundary flux correction:
\begin{equation}
F_q^c-F_q^v
=\mathcal F^c(U)\cdot\hat n-b(U).
\label{eq:robin_total_boundary_flux}
\end{equation}

\section[Hybrid Boundary-Layer Meshes and NURBS-Enhanced Curved Geometry]
{Hybrid Boundary-Layer Meshes and NURBS-Enhanced Curved Geometry}
\label{sec:curved_geometry}

\subsection{Hybrid triangular--quadrilateral mesh strategy}

Laminar boundary layers exhibit disparate length scales: variation across the thin layer is generally much sharper than along the wall. Resolving the smallest wall-normal scale with nearly isotropic cells would impose unnecessary tangential refinement and increase 
the computational cost.
Anisotropic near-wall meshes instead concentrate mesh resolution
across the dominant gradients, improving the numerical efficiency of viscous-flow
calculations~\cite{venditti2003,moro2017}.

In our hybrid mesh, stretched quadrilateral layers approximately follow
the local wall-tangential and wall-normal directions, allowing normal
spacing and its growth through the layer to be controlled largely
independently of tangential spacing. Farther out, as the strong anisotropy in the wall-normal direction weakens, we use an unstructured triangular mesh to provide flexible transitions
around complex geometry and through far-field and wake regions without
extending the stretched layers throughout the domain. Dussin
\emph{et al.}~\cite{dussin2009} illustrate this hybrid-grid rationale
for compressible viscous flows.

The large aspect ratios of the near-wall quadrilateral
cells complicate accurate least-squares reconstruction near curved walls.
Jalali and Ollivier-Gooch~\cite{jalali2017rans} showed that high-order
$k$-exact least-squares reconstruction loses accuracy on highly anisotropic
meshes in the presence of curvature, and improved the reconstruction using
local curvilinear tangential--normal coordinates.
Huang \emph{et al.}~\cite{huang2022mixed} optimized variational
reconstruction to reduce reconstruction errors on quadrilateral grids
with large aspect ratios.
In the present method, both quadrilateral and triangular
cells use the $k$-exact reconstruction of conservative variables described in
Section~\ref{sec:kexact}, with geometric moments evaluated over the
corresponding control volumes. At interfaces between the two cell types,
the numerical fluxes are computed from the reconstructed states and
gradients on either side using the interior-face formulation in
Section~\ref{sec:viscous_flux}.

Another challenge is how to accurately handle the curved wall boundary,
since the highly stretched quadrilateral cells are
particularly sensitive to geometric approximation errors.
The polygonal approximation of a curved wall boundary introduces geometric errors into numerical quadrature and may thereby produce nonphysical numerical solutions in the simulation of inviscid compressible flows~\cite{bassi1997,menghu2018}. We therefore employ NURBS curves
to represent the curved wall boundary and provide accurate
curved-boundary quadrature for numerical flux integration
and curved-cell moments for the $k$-exact reconstruction,
as presented in Sections~4.2--4.3.

\subsection{NURBS curve and boundary integration}

Following Meng and Hu~\cite{menghu2018}, a curved wall boundary is parameterized by a NURBS curve
\begingroup
\predisplaypenalty=0
\begin{equation}
\vec C(\xi)
= \frac{\sum_{j=1}^{n}N_{j,p_g}(\xi)w_j\vec P_j}
{\sum_{j=1}^{n}N_{j,p_g}(\xi)w_j},
\qquad \xi\in[0,1],
\label{eq:nurbs_def}
\end{equation}
\endgroup
where $\vec P_j$ are control points, $w_j>0$ are weights, $n$ is the number of B-spline basis functions, and
$N_{j,p_g}$ are degree-$p_g$ B-spline basis functions, which are defined on the knot vector
$\Xi=\{\xi_1,\ldots,\xi_{n+p_g+1}\}$~\cite{piegl1997}. When the curved wall boundary cannot be represented exactly by a NURBS curve, we construct a curve through the boundary mesh points using global interpolation with chord-length parameterization and weights $w_j=1$; see~\cite{menghu2018} for details.

A curved edge $e$ is a trimmed NURBS curve $\vec C(\xi)$ with
$\xi\in[\xi_a,\xi_b]\subseteq[0,1]$. Its line integral is
\begin{equation}
\int_e g(\vec x)\,\mathrm ds
=\int_{\xi_a}^{\xi_b}g(\vec C(\xi))
\|\vec C'(\xi)\|_2\,\mathrm d\xi,
\label{eq:curved_face_integral}
\end{equation}
and we use the Gauss--Legendre quadrature rule to compute it.

The quadrature points, outward normals, and weights obtained
from Eq.~\eqref{eq:curved_face_integral} are used to evaluate the boundary
fluxes in Eq.~\eqref{eq:discrete_residual}. To maintain geometric consistency
in the discrete flux integration, the quadrature weights are corrected to
match the edge length $L_e\approx\int_e\mathrm ds$ and integrated normal
$\vec N_e\approx\int_e\hat n\,\mathrm ds$, evaluated by Gauss quadrature on
each knot subinterval. The conditions $\sum_q w_q=L_e$ and
$\sum_q w_q\hat n_q=\vec N_e$ give
\begin{equation}
\begin{gathered}
A_e=\begin{bmatrix}
1&\cdots&1\\
\hat n_{1,x}&\cdots&\hat n_{N_e,x}\\
\hat n_{1,y}&\cdots&\hat n_{N_e,y}
\end{bmatrix},\qquad
b_e=\begin{pmatrix}L_e\\N_{e,x}\\N_{e,y}\end{pmatrix},\\
\delta w_e=A_{e,\tau_e}^{+}
(b_e-A_e w_e^{(0)}).
\end{gathered}
\label{eq:metric_correction}
\end{equation}
Here $w_e=w_e^{(0)}+\delta w_e$, and
$A_{e,\tau_e}^+$ is the pseudoinverse computed by singular value decomposition (SVD) with cutoff $\tau_e$.
The correction is applied when the moment conditions are satisfied to roundoff
precision and all weights remain positive.

\subsection{Numerical integration over curved cells}

We evaluate volume integrals over the curved quadrilateral cells in the thin near-wall region. Following~\cite{sevilla2008,menghu2018}, for a cell $\Omega_i$ with one curved edge $e=\vec C[\xi_a,\xi_b]$, the geometric mapping from $[0,1]^2$ to the physical cell $\Omega_i$ is given by
\begin{equation}
\vec X_i(\xi,\lambda)
=(1-\lambda)\vec C\bigl(\xi_a+(\xi_b-\xi_a)\xi\bigr)
+\lambda \big( (1-\xi) \vec x_1^{(i)} + \xi \vec x_2^{(i)}\big),
\label{eq:cell_part}
\end{equation}
where $(\xi,\lambda)\in[0,1]^2$, and $\vec x_1^{(i)}$, $\vec x_2^{(i)}$ are the two interior vertices of $\Omega_i$. Then we have
\begin{equation}
\int_{\Omega_i}g(\vec x)\,\mathrm d\Omega
=\int_0^1\!\int_0^1 g(\vec X_i(\xi,\lambda))
\left|\det D\vec X_i\right|\,\mathrm d\xi\,\mathrm d\lambda,
\label{eq:curved_cell_integral}
\end{equation}
where $D\vec X_i=\partial\vec X_i/\partial(\xi,\lambda)$, and we use the tensor-product Gauss--Legendre quadrature rule to evaluate the integral.

We therefore evaluate the curved-cell area $A_i$ and the
scaled control-volume moments $\overline{\psi_{i,\alpha}}_i$ using the
quadrature in Eq.~\eqref{eq:curved_cell_integral}. These moments enter the
mean-preserving reconstruction basis in Eq.~\eqref{eq:monomials} and, when
curved cells belong to a reconstruction stencil, the cell-averaged basis
functions used in Eq.~\eqref{eq:kexact_constraint}. The resulting reconstruction provides
the states and gradients for the numerical fluxes, while the curved-cell
area enters the finite-volume residual in Eq.~\eqref{eq:discrete_residual}.


\section{Preconditioned Matrix-Free Newton--Krylov Solver}

The method used to solve the nonlinear system~\eqref{eq:ode} follows the approach in Section~3 of Nejat and
Ollivier-Gooch~\cite{nejat2008preconditioning}, where the
inviscid Euler equations are considered. In this paper, a thin-layer approximation to the viscous flux Jacobian is additionally assembled as part of the preconditioning matrix.

On a fixed mesh, the steady solution is obtained by
pseudo-time evolution~\cite{kelley1998}:
\begin{equation}
\frac{\mathrm d\overline U_i}{\mathrm d\tau}
=-R_i(\overline U),
\label{eq:pseudo_time_evolution}
\end{equation}
where $R_i$ is defined by Eq.~\eqref{eq:discrete_residual}.
The backward Euler scheme gives
$\delta U_i/\Delta t + R_i(\overline U^n+\delta U)=0$.
Linearizing the residual about $\overline U^n$ yields
\begin{equation}
\begin{aligned}
\left(\frac{I}{\Delta t}+J^n\right)\delta U
&=-R(\overline U^n),\\
J^n=\left.\frac{\partial R}{\partial\overline U}\right|_{\overline U^n},
\qquad
\delta U&=\overline U^{n+1}-\overline U^n.
\end{aligned}
\label{eq:nejat2008-13}
\end{equation}
When $\Delta t\to\infty$, this system reduces to Newton's method.
\subsection{Linear solver and preconditioning}
\label{sec:matrix_free_solver}
\label{sec:solver_preconditioning}

The generalized minimal residual method (GMRES) solves the linearized system without explicitly forming the
high-order Jacobian. The required matrix--vector products are approximated
by directional finite differences~\cite{knoll2004}:
\begin{equation}
J^nz\approx
\frac{R(\overline U^n+\varepsilon z)-R(\overline U^n-\varepsilon z)}{2\varepsilon},
\qquad
\varepsilon=\frac{\varepsilon_0\sqrt{1+\|\overline U^n\|_2}}{\|z\|_2}.
\label{eq:nejat2008-15}
\end{equation}
The default reference perturbation parameter is $\varepsilon_0=10^{-8}$.

To precondition the linear system, let $A=I/\Delta t+J^n$
and $b=-R(\overline U^n)$.
Preconditioning gives
\begin{equation}
A M^{-1}y=b,\qquad \delta U=M^{-1}y,
\label{eq:right_preconditioning}
\end{equation}
where $M$ is a sparse, low-order approximation to $A$.
The system involving $M$ is solved by LU or incomplete LU factorization,
ILU($p$). Reverse Cuthill--McKee reordering reduces fill-in, while
increasing $p$ improves the factorization at additional storage and
computational cost.

\subsection{Preconditioner matrix}
\label{sec:preconditioner_matrix}

The first-order Jacobian couples $\Omega_i$ only to cells sharing a face.
Let $\mathcal N_i$ denote their indices; interior triangular and
quadrilateral cells have three and four such neighbors, respectively.
For the common face $e_{ij}$, take $U_L$ from $\Omega_i$, $U_R$ from
$\Omega_j$, and the normal pointing outward from $\Omega_i$.
Let $B_{e,q}^L$ and $B_{e,q}^R$ be the approximate $4\times4$
derivatives of $F_q^c-F_q^v$ with respect to the two adjacent cell
averages. The nonzero preconditioning blocks are
\begin{equation}
\begin{aligned}
M_{ii}&=\frac{I_4}{\Delta t}
+\frac{1}{A_i}\sum_{e\subset\partial\Omega_i}
\sum_{q=1}^{N_e}w_q B_{e,q}^L,\\
M_{ij}&=\frac{1}{A_i}\sum_{q=1}^{N_{e_{ij}}}
w_q B_{e_{ij},q}^R,\qquad j\in\mathcal N_i.
\end{aligned}
\label{eq:compact_preconditioner_blocks}
\end{equation}

For the approximate analytical convective Jacobian, the Roe dissipation
matrix $D=|\widetilde A_n|$ is held fixed:
\begin{equation}
B^{c,L}=\tfrac12\bigl(A_n(U_L)+D\bigr),\qquad
B^{c,R}=\tfrac12\bigl(A_n(U_R)-D\bigr),
\label{eq:approximate_roe_jacobian}
\end{equation}
where $A_n(U)=\partial(\mathcal F^c\cdot\hat n)/\partial U$.
As in Section~3.3.1 of~\cite{nejat2008preconditioning}, the Jacobian
coefficients are evaluated using linearly reconstructed quadrature-point
states, with reconstruction corrections held fixed.
Alternatively, the convective blocks are obtained by finite differences
of the face fluxes.

The viscous blocks differentiate the interface velocity and transport
coefficients and use a two-point thin-layer approximation for gradient
derivatives. With $V=(\rho,u,v,T)^T$,
\begin{equation}
(\nabla V)_{\rm TL}
=\bigl(V(U_R)-V(U_L)\bigr)\otimes
\frac{\vec d_{ij}}{\|\vec d_{ij}\|_2^2},
\qquad \vec d_{ij}=\vec x_j-\vec x_i.
\label{eq:viscous_thin_layer}
\end{equation}
The gradient value remains reconstructed; only its derivative is
approximated. On interior dGRP faces, the state-jump term in
Eq.~\eqref{eq:dgrp} and the conversion to velocity and temperature
gradients are differentiated with the one-sided conservative gradients
fixed. Transport-coefficient derivatives vanish for constant viscosity.
At physical boundaries, exterior-state derivatives contribute only to the
adjacent interior-cell diagonal block; at adiabatic walls, the normal
temperature-gradient component and its derivative are removed.

\subsection{Start-up and Newton iterations}
\label{sec:solver_startup}
\label{sec:solver_newton}

An implicit defect-correction iteration provides the initial state for
Newton iteration:
\begin{equation}
\left(\frac{I}{\Delta t}+J_1^n\right)\delta U
=-R(\overline U^n),
\label{eq:defect_correction_startup}
\end{equation}
where $J_1^n$ is the first-order Jacobian approximation and the residual
retains the high-order discretization. The approximate analytical
Jacobian and ILU(1) are used in this phase, with linear residual
tolerance $5\times10^{-2}\|R(\overline U^n)\|_2$.

Once an initial state is available,
the high-order pseudo-transient Newton system
\begin{equation}
\left(\frac{I}{\Delta t^n}+J^n\right)\delta U=-R(\overline U^n)
\label{eq:newton_phase}
\end{equation}
is solved by matrix-free GMRES, preconditioned by the low-order
matrix with finite-difference convective Jacobian blocks and ILU($p$).
Both phases use restarted GMRES with a maximum of $200$ search directions
per restart cycle.

\section{Numerical Results}
\label{sec:numerical_results}
\makeatletter
\setlength{\@dblfpsep}{4pt plus 1fil}
\makeatother

In this section, we present two numerical examples on curved domains to assess spatial accuracy and three applications involving
compressible laminar boundary-layer flows. Table~\ref{tab:case_configuration}
summarizes the configurations. All numerical examples use unlimited $k$-exact reconstruction
with the numerical convective and viscous fluxes and boundary treatment techniques developed in Sections~3 and~4. For the accuracy
tests, we set the nondimensional gas constant to $R=1/(\gamma M^2)$.

For the NACA0012, tandem-airfoil, and Gaussian-bump cases, initial
approximations for the single airfoil and the coarsest grid of each mesh sequence are
obtained from uniform freestream conditions using the start-up procedure in
Section~\ref{sec:solver_startup}. The single-airfoil and coarse tandem calculations each
use $10$ start-up updates, whereas a relative energy-residual tolerance of $10^{-2}$ is
used for the bump start-up. Finer grids are initialized by interpolating the converged
solution on the preceding grid. The subsequent Newton--Krylov iterations use GMRES(200)
with ILU(8) preconditioning and a relative linear tolerance of $10^{-5}$.
The initial Courant--Friedrichs--Lewy (CFL) number is $100$ for NACA0012 and $1000$ for the tandem and bump cases, with an upper
bound of $10^6$. The relative energy-residual tolerances are $10^{-10}$ for the airfoil
cases and $10^{-6}$ for the Gaussian bump.

Spatial accuracy is measured by the $L^2$ error in the velocity field $\vec{v}$:
\begin{equation}
\epsilon_v
=\left[\sum_i\int_{\Omega_i}
\|\vec v_h-\vec v\|_2^2\,\mathrm d\Omega\right]^{1/2},
\label{eq:rms_errors}
\end{equation}
where $\vec v_h=((\rho u)_h,(\rho v)_h)^T/\rho_h$ is the numerical velocity
and $\vec v$ is the exact velocity. The characteristic mesh size is $h=\sqrt{|\Omega|/N}$.

\begin{table*}[!tbp]
\centering
\caption{Numerical configurations. $M$ is the inner-wall Mach number
for Couette, the reference Mach number for MMS, and the freestream
Mach number for the external flows.}
\label{tab:case_configuration}
\begingroup
\small
\setlength{\tabcolsep}{5pt}
\renewcommand{\arraystretch}{1.2}
\begin{tabularx}{\textwidth}{@{}
  >{\hsize=0.9\hsize\linewidth=\hsize\raggedright\arraybackslash}X
  cc
  >{\hsize=1.2\hsize\linewidth=\hsize\raggedright\arraybackslash}X
  cc
  >{\hsize=0.9\hsize\linewidth=\hsize\raggedright\arraybackslash}X@{}}
\toprule
Case & $M$ & $\mathrm{Re}$ & Wall condition & $k$ & Flux & Mesh (cells) \\
\midrule
Couette & 0.5 & 10 & Inner: isothermal;\newline outer: adiabatic
 & 1, 2, 3 & Roe & $646$--$41{,}344$ \\
Elliptical MMS & 0.5 & 100 & Isothermal
 & 1, 2, 3 & Roe & $646$--$41{,}344$ \\
NACA0012 & 0.5 & 5{,}000 & Adiabatic
 & 3 & Roe & $35{,}292$ \\
Tandem NACA0012 & 0.5 & 5{,}000 & Adiabatic
 & 3 & Roe & $13{,}178$--$123{,}404$ \\
Gaussian bump & 0.5 & $4\times10^5$ & Isothermal
 & 3 & HLLC & $25{,}391$--$128{,}226$ \\
\bottomrule
\end{tabularx}
\endgroup
\end{table*}

\begin{table*}[tbp]
\centering
\caption{Start-up and main Newton--Krylov updates. The single-airfoil case and the coarsest grid of each sequence start from uniform freestream conditions; finer grids use interpolation from the converged solution on the preceding grid in the same sequence. Totals include only updates on the indicated grid; updates on preceding grids are reported in their respective rows.}
\label{tab:application_iterations}
\begin{tabular}{@{}lrrrr@{}}
\toprule
Case & $N$ & Start-up & Main & Total \\
\midrule
NACA0012 & 35{,}292 & 10 & 146 & 156 \\
Tandem NACA0012 & 13{,}178 & 10 & 21 & 31 \\
Tandem NACA0012 & 57{,}624 & 0 & 18 & 18 \\
Tandem NACA0012 & 123{,}404 & 0 & 13 & 13 \\
Gaussian bump & 25{,}391 & 78 & 29 & 107 \\
Gaussian bump & 56{,}992 & 0 & 16 & 16 \\
Gaussian bump & 128{,}226 & 0 & 32 & 32 \\
\bottomrule
\end{tabular}
\end{table*}

\subsection{Cylindrical Couette Flow}
\label{sec:couette}

In this example, we consider steady cylindrical
Couette flow between two concentric cylinders, following Michalak and
Ollivier-Gooch~\cite{michalak2009}. The inner and outer radii are $r_i=1$
and $r_o=2$, respectively. The inner cylinder rotates clockwise with unit
tangential speed, while the outer cylinder is stationary and adiabatic.
We use an inner-wall Mach number of $M=0.5$ and a Reynolds number of
$\mathrm{Re}=10$. The Reynolds number is defined using the inner-cylinder
radius $r_i$ as the characteristic length and the inner-wall tangential
speed as the characteristic velocity. For the inner cylinder, 
we impose the isothermal boundary condition $T=T_i$.

With constant viscosity and thermal conductivity, the tangential-speed
magnitude is
\begin{equation}
v_\theta(r)=\frac{r_i}{r}\frac{r_o^2-r^2}{r_o^2-r_i^2},
\label{eq:couette_solution}
\end{equation}
with Cartesian velocity $\vec v=v_\theta(y/r,-x/r)^T$.
The temperature is
\begin{equation}
\begin{gathered}
T(r)=1+\Theta C^2
\left[\frac14-\frac{1}{r^2}-\frac12\log\left(\frac r2\right)\right],
\\ \Theta=M^2(\gamma-1)\mathrm{Pr},\quad C=\frac43.
\end{gathered}
\label{eq:couette_temperature_reference}
\end{equation}
The normalization $T(2)=1$ determines $T_i=T(1)$.
Pressure and density follow from radial momentum balance,
\begin{equation}
\frac{\mathrm dP}{\mathrm dr}=\frac{\rho v_\theta^2}{r},
\label{eq:couette_pressure_balance}
\end{equation}
with $P(1)=R$ and the equation of state. 

\begin{figure*}[t]
\centering
\includegraphics[width=0.66\textwidth]{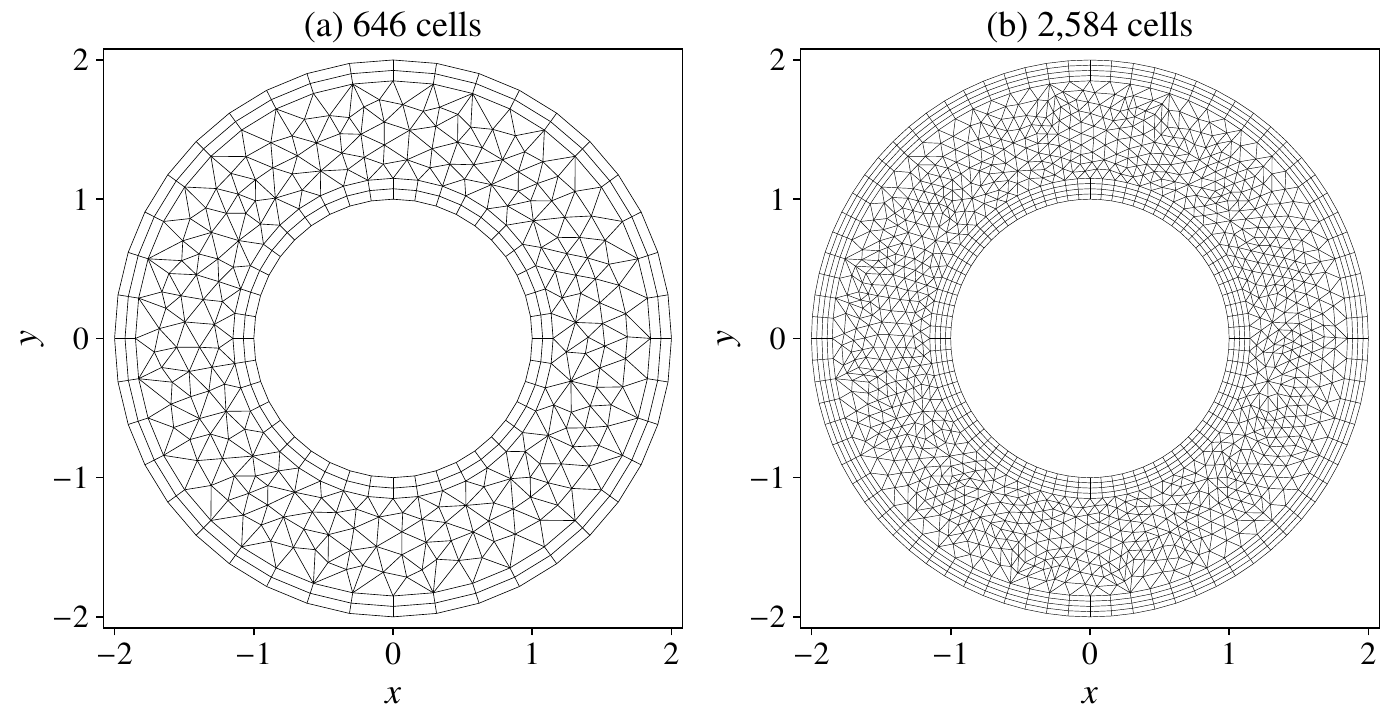}
\caption{The first two hybrid meshes for the simulation of cylindrical Couette flow.}
\label{fig:couette_hybrid_grids}
\end{figure*}

We use a nested hierarchy of hybrid meshes, with each finer mesh
obtained by uniform refinement of the preceding mesh. We show the
first two levels in Figure~\ref{fig:couette_hybrid_grids}, and present
the $L^2$ errors for velocity in
Table~\ref{tab:couette_hybrid_conv}.
The numerical results are obtained using the $k$-exact NURBS-enhanced finite-volume method with $1\le k \le 3$.
For comparison, we also report results obtained using a polygonal approximation of the curved boundaries.
With NURBS geometry, the nominal convergence order is achieved except for $k=2$,
and the velocity errors are generally smaller than those obtained with polygonal geometry.
These results demonstrate the effectiveness of the NURBS representation.

\begin{table*}[!t]
\centering
\caption{Cylindrical Couette flow: convergence of $L^2$ errors in velocity.}
\label{tab:couette_hybrid_conv}
\begingroup
\small
\setlength{\tabcolsep}{5pt}
\begin{tabular}{@{}ccrcccccc@{}}
\toprule
 & & & \multicolumn{2}{c}{Polygon} & \multicolumn{2}{c}{NURBS} & \multicolumn{2}{c}{NURBS, without dGRP} \\
\cmidrule(lr){4-5}\cmidrule(lr){6-7}\cmidrule(l){8-9}
$k$ & Level & $N$ & $\epsilon_v$ & Order & $\epsilon_v$ & Order & $\epsilon_v$ & Order \\
\midrule
\multirow{4}{*}{1} & 1 & 646 & $1.100\times10^{-2}$ & --- & $8.223\times10^{-3}$ & --- & not converged & --- \\
 & 2 & 2{,}584 & $2.727\times10^{-3}$ & 2.01 & $2.020\times10^{-3}$ & 2.03 & not converged & --- \\
 & 3 & 10{,}336 & $6.653\times10^{-4}$ & 2.04 & $4.884\times10^{-4}$ & 2.05 & not converged & --- \\
 & 4 & 41{,}344 & $1.642\times10^{-4}$ & 2.02 & $1.200\times10^{-4}$ & 2.03 & not converged & --- \\
\midrule
\multirow{4}{*}{2} & 1 & 646 & $5.294\times10^{-3}$ & --- & $2.780\times10^{-3}$ & --- & $2.270\times10^{-3}$ & --- \\
 & 2 & 2{,}584 & $1.334\times10^{-3}$ & 1.99 & $7.565\times10^{-4}$ & 1.88 & $7.832\times10^{-4}$ & 1.53 \\
 & 3 & 10{,}336 & $3.174\times10^{-4}$ & 2.07 & $1.713\times10^{-4}$ & 2.14 & $1.827\times10^{-4}$ & 2.10 \\
 & 4 & 41{,}344 & $7.654\times10^{-5}$ & 2.05 & $3.954\times10^{-5}$ & 2.11 & $4.234\times10^{-5}$ & 2.11 \\
\midrule
\multirow{4}{*}{3} & 1 & 646 & $5.085\times10^{-3}$ & --- & $2.297\times10^{-3}$ & --- & $1.998\times10^{-3}$ & --- \\
 & 2 & 2{,}584 & $7.954\times10^{-4}$ & 2.68 & $8.289\times10^{-5}$ & 4.79 & $9.514\times10^{-5}$ & 4.39 \\
 & 3 & 10{,}336 & $1.884\times10^{-4}$ & 2.08 & $5.534\times10^{-6}$ & 3.90 & $4.376\times10^{-6}$ & 4.44 \\
 & 4 & 41{,}344 & $4.626\times10^{-5}$ & 2.03 & $3.493\times10^{-7}$ & 3.99 & $2.631\times10^{-7}$ & 4.06 \\
\bottomrule
\end{tabular}
\endgroup
\end{table*}

\begingroup
We also compute the Couette flow using the averaged-gradient viscous
flux described in Section~\ref{sec:viscous_flux}. The results are included
in Table~\ref{tab:couette_hybrid_conv} alongside the Polygon and NURBS
results with dGRP. The NURBS geometry, reconstruction, and wall treatment
are the same in both calculations. The averaged-gradient calculations
do not converge for linear reconstruction.
For cubic reconstruction, both fluxes
give approximately fourth-order convergence in velocity, with comparable
errors on the finest mesh.
\par
\endgroup

\subsection{Elliptical Manufactured Solution}
\label{sec:elliptic_mms}

In the second numerical example, we construct a manufactured solution for a nonaxisymmetric flow with variable transport coefficients in the elliptical annulus
\begin{equation}
\begin{gathered}
 \Omega=\{(x,y):1\le s\le4\},
\\
 s=\left(\frac{x}{a_x}\right)^2+\left(\frac{y}{a_y}\right)^2,
 \qquad a_x=\frac54,\quad a_y=1.
\end{gathered}
\label{eq:elliptic_mms_domain}
\end{equation}
We use $M=0.5$ and $\mathrm{Re}=100$. No-slip isothermal boundary conditions with $T_w=1$ are imposed on both ellipses. The reference fields are
\begin{equation}
\begin{aligned}
 B&=\frac49(s-1)(4-s),\\
 \psi&=0.1B^2(1+0.2x-0.15y),\\
 \rho &=1+0.1\sin x+0.05\sin y+0.02xy,\\
 \rho \vec v
 &=(\partial_y\psi,-\partial_x\psi)^T,\\
 T &=1+0.1B(1+0.2x-0.1y).
\end{aligned}
\label{eq:elliptic_mms_fields}
\end{equation}
Pressure and total energy follow from Eq.~\eqref{eq:eos}, and the viscosity
follows Sutherland's law. We consider the compressible
Navier--Stokes equations with a source term:
\begin{equation}
\nabla\cdot\left[
\mathcal F^c(U)
-\mathcal F^v(U,\nabla U)\right]=S.
\label{eq:elliptic_mms_source}
\end{equation}
The source vector $S$ is computed by substituting the analytical solution
in Eq.~\eqref{eq:elliptic_mms_fields} into the flux divergence on the
left-hand side. The construction
$\rho\vec v=(\partial_y\psi,-\partial_x\psi)^T$ guarantees
$\nabla\cdot(\rho\vec v)=0$, so the component of $S$ corresponding to
mass conservation vanishes identically, i.e., $S_\rho=0$. Thus, the
continuity equation has no additional mass source term.

\begin{figure*}[tbp]
\centering
\includegraphics[width=0.95\textwidth]{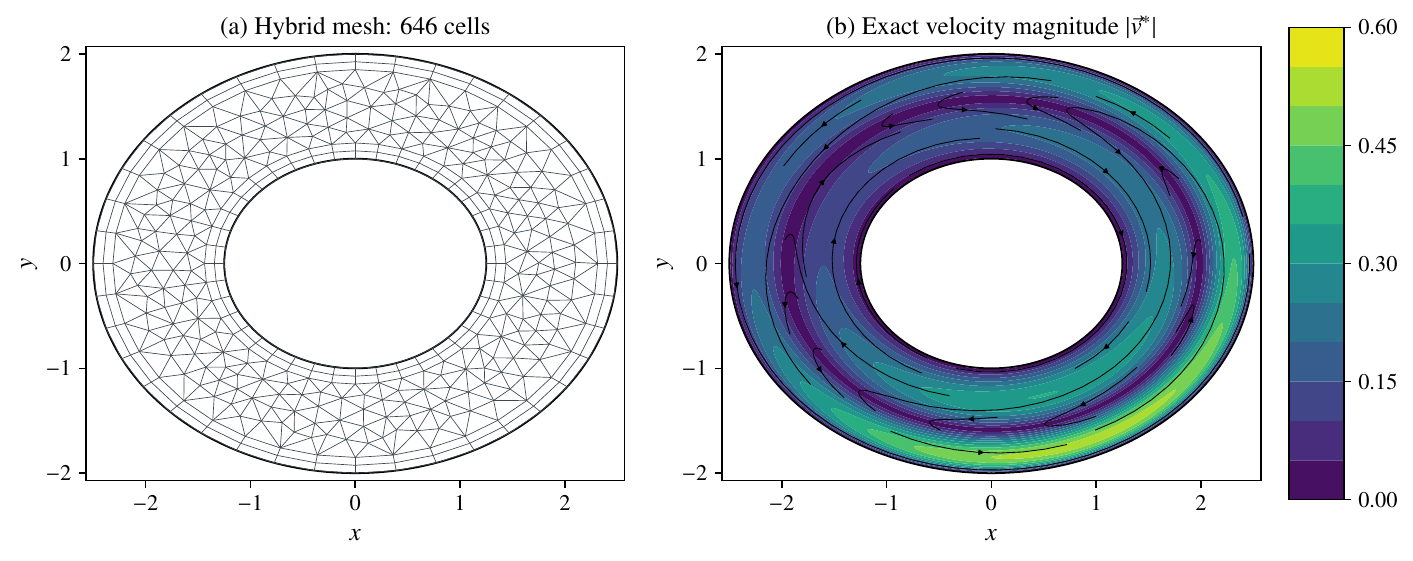}
\caption{(a) The coarsest hybrid NURBS mesh;
(b)~exact velocity magnitude and streamlines.}
\label{fig:elliptic_mms_reference}
\end{figure*}

As in the previous example, the boundaries can be represented exactly by NURBS curves.
In Figure~\ref{fig:elliptic_mms_reference}, we show the coarsest mesh, the exact
velocity magnitude, and streamlines. We present the $L^2$ errors in velocity in Table~\ref{tab:elliptic_mms_conv}.
The numerical results are obtained using the $k$-exact NURBS-enhanced finite-volume method with $1\le k \le 3$.
For comparison, we also report results obtained using a polygonal approximation of the curved boundaries.
With NURBS geometry, the nominal convergence order in velocity is achieved except for $k=2$.
For cubic reconstruction, the NURBS errors on the finer meshes are much smaller than the polygonal errors,
again demonstrating the effectiveness of the NURBS representation.

\begin{table*}[tbp]
\centering
\caption{Elliptical manufactured solution: convergence of $L^2$ errors in velocity.}
\label{tab:elliptic_mms_conv}
\begingroup
\small
\setlength{\tabcolsep}{5pt}
\begin{tabular}{@{}ccrcccccc@{}}
\toprule
 & & & \multicolumn{2}{c}{Polygon} & \multicolumn{2}{c}{NURBS} & \multicolumn{2}{c}{NURBS, without dGRP} \\
\cmidrule(lr){4-5}\cmidrule(lr){6-7}\cmidrule(l){8-9}
$k$ & Level & $N$ & $\epsilon_v$ & Order & $\epsilon_v$ & Order & $\epsilon_v$ & Order \\
\midrule
\multirow{4}{*}{1} & 1 & 646 & $2.386\times10^{-1}$ & --- & $2.529\times10^{-1}$ & --- & not converged & --- \\
 & 2 & 2{,}584 & $6.367\times10^{-2}$ & 1.91 & $6.773\times10^{-2}$ & 1.90 & not converged & --- \\
 & 3 & 10{,}336 & $1.557\times10^{-2}$ & 2.03 & $1.655\times10^{-2}$ & 2.03 & not converged & --- \\
 & 4 & 41{,}344 & $3.840\times10^{-3}$ & 2.02 & $4.081\times10^{-3}$ & 2.02 & not converged & --- \\
\midrule
\multirow{4}{*}{2} & 1 & 646 & $2.304\times10^{-2}$ & --- & $2.971\times10^{-2}$ & --- & $5.794\times10^{-2}$ & --- \\
 & 2 & 2{,}584 & $5.559\times10^{-3}$ & 2.06 & $3.811\times10^{-3}$ & 2.96 & $1.895\times10^{-2}$ & 1.61 \\
 & 3 & 10{,}336 & $1.570\times10^{-3}$ & 1.83 & $7.601\times10^{-4}$ & 2.33 & $5.141\times10^{-3}$ & 1.88 \\
 & 4 & 41{,}344 & $4.768\times10^{-4}$ & 1.72 & $2.527\times10^{-4}$ & 1.59 & $1.355\times10^{-3}$ & 1.92 \\
\midrule
\multirow{4}{*}{3} & 1 & 646 & $3.023\times10^{-2}$ & --- & $4.325\times10^{-2}$ & --- & $4.390\times10^{-2}$ & --- \\
 & 2 & 2{,}584 & $3.388\times10^{-3}$ & 3.16 & $1.070\times10^{-3}$ & 5.34 & $3.570\times10^{-3}$ & 3.62 \\
 & 3 & 10{,}336 & $9.511\times10^{-4}$ & 1.83 & $5.756\times10^{-5}$ & 4.22 & $2.386\times10^{-4}$ & 3.90 \\
 & 4 & 41{,}344 & $2.419\times10^{-4}$ & 1.98 & $3.483\times10^{-6}$ & 4.05 & $1.729\times10^{-5}$ & 3.79 \\
\bottomrule
\end{tabular}
\endgroup
\end{table*}

\begingroup
The averaged-gradient results for this manufactured solution are also
included in Table~\ref{tab:elliptic_mms_conv}, using the same NURBS
geometry, reconstruction, and wall treatment. The Polygon and NURBS
columns use dGRP. For cubic reconstruction,
the finest-grid convergence orders for velocity are $4.05$ with dGRP and $3.79$ with
averaged gradients. Both fluxes converge for quadratic and cubic
reconstruction, while the averaged-gradient calculations do not converge
for linear reconstruction.
\par
\endgroup

\begingroup
\noindent\begin{minipage}{\linewidth}
\begin{remark}[Odd--even behavior of viscous convergence]
\label{rem:quadratic_convergence}
For nominal order $p=k+1$, leading gradient errors on locally symmetric
stencils have opposite signs for even $p$ and can cancel in the interface
average. For odd $p$, they generally reinforce and may lower viscous
accuracy by one order. Ollivier-Gooch and
Van~Altena~\cite{olliviergooch2002} predict diffusion orders $2$, $2$, and
$4$ for $k=1,2,3$, consistent with the trends in both tests.
\end{remark}

\begin{remark}[Viscous coupling with linear reconstruction]
\label{rem:linear_viscous_coupling}
For $k=1$, averaging the cellwise constant gradients can leave
insufficient intercell coupling~\cite{jalali2014diffusive}. In both tests,
the averaged-gradient flux fails to meet the nonlinear stopping criterion
on all four meshes, whereas dGRP converges. With the same wall treatment,
a two-point centerline gradient correction restores convergence and
yields finest-grid convergence orders for velocity of $2.02$ (Couette) and $1.98$
(elliptical MMS).
\end{remark}
\end{minipage}\par
\endgroup


\subsection{Laminar Boundary Layer over a NACA0012 Airfoil}
\label{sec:naca}

\begin{figure*}[!tbp]
\centering
\includegraphics[width=0.95\textwidth]{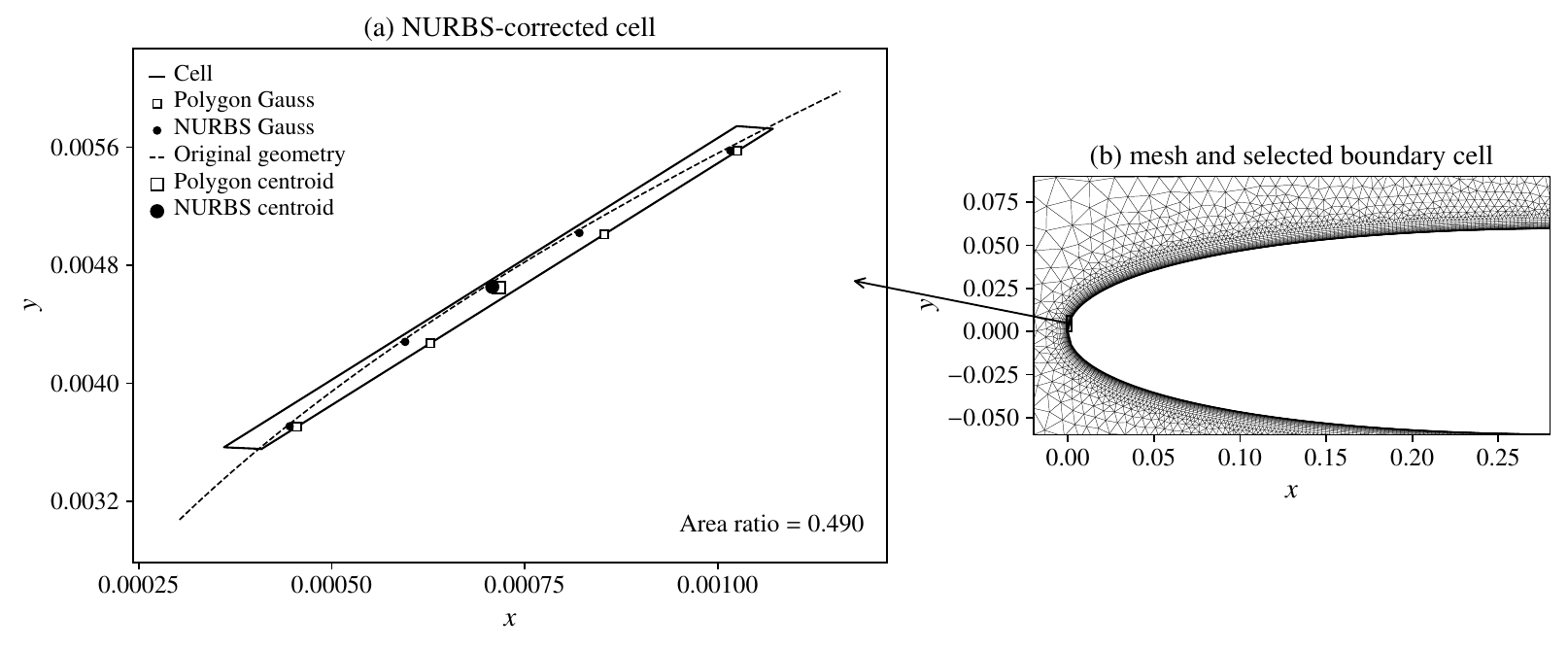}
\caption{NACA0012: (a)~polygonal and NURBS representations of a leading-edge
boundary cell; (b)~local hybrid mesh. The area ratio compares the curved
cell with its straight-sided counterpart.}
\label{fig:naca_mesh_nurbs}
\end{figure*}

We next consider subsonic laminar compressible flow around the NACA0012 airfoil at zero angle of attack, with freestream Mach number $M=0.5$ and Reynolds number $\mathrm{Re}=5000$ based on the chord length.
No-slip adiabatic boundary conditions are imposed on the airfoil, with
inflow/outflow boundary conditions at the outer boundary.
Figure~\ref{fig:naca_mesh_nurbs} shows the hybrid
mesh near the leading edge and compares polygonal and NURBS
representations of a boundary cell.

For the three external-flow applications, the surface coefficients are
\begin{equation}
C_p=\frac{P-P_\infty}{q_\infty},\qquad
C_f=\frac{\tau_w}{q_\infty},\qquad
\tau_w=(\tau\hat n)\cdot\hat t,
\label{eq:wall_coefficients}
\end{equation}
where $q_\infty=\tfrac12\rho_\infty U_\infty^2$,
$U_\infty=\|\vec v_\infty\|_2$, and $\hat n$ is the outward
fluid normal. The tangent $\hat t=(\hat n_y,-\hat n_x)^T$ sets the sign of $C_f$;
surface values are wall-face arc-length averages. The drag coefficients are
\begin{equation}
\begin{aligned}
C_{D,\mathrm p}
&=\frac{1}{q_\infty L_{\mathrm{ref}}}
\int_{S_w}(P-P_\infty)\,\hat n\cdot
\hat e_D\,\mathrm ds,\\
C_{D,\mathrm v}
&=-\frac{1}{q_\infty L_{\mathrm{ref}}}
\int_{S_w}(\tau\hat n)\cdot
\hat e_D\,\mathrm ds,
\qquad C_D=C_{D,\mathrm p}+C_{D,\mathrm v}.
\end{aligned}
\label{eq:drag_convention}
\end{equation}
Here $S_w$ is the body surface, $\hat e_D$ the freestream
direction, and $L_{\mathrm{ref}}$ the airfoil chord or unit length for the
Gaussian bump. Lift is positive in $+y$; moments are positive
counterclockwise and normalized by $q_\infty L_{\mathrm{ref}}^2$,
with centers given in the tables.

\begin{figure}[!tbp]
\centering
\includegraphics[width=0.95\textwidth]{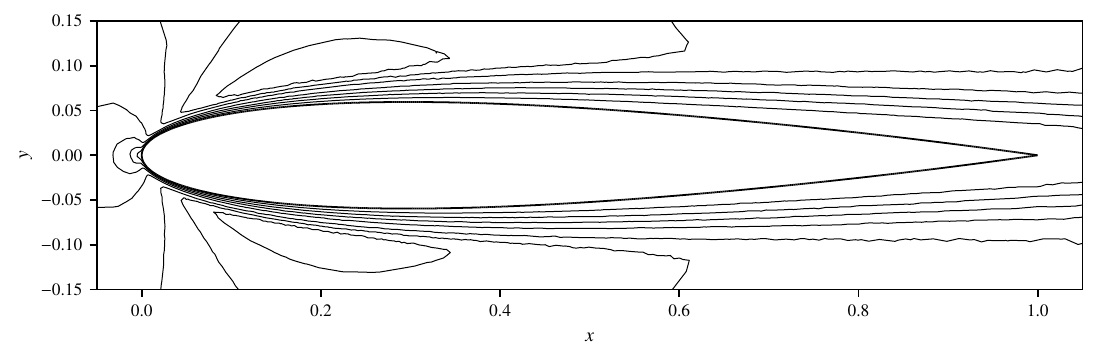}
\caption{NACA0012: Mach-number contours showing the boundary layer and wake.}
\label{fig:naca_mach}
\end{figure}

\begin{figure*}[!tbp]
\centering
\includegraphics[width=0.95\textwidth]{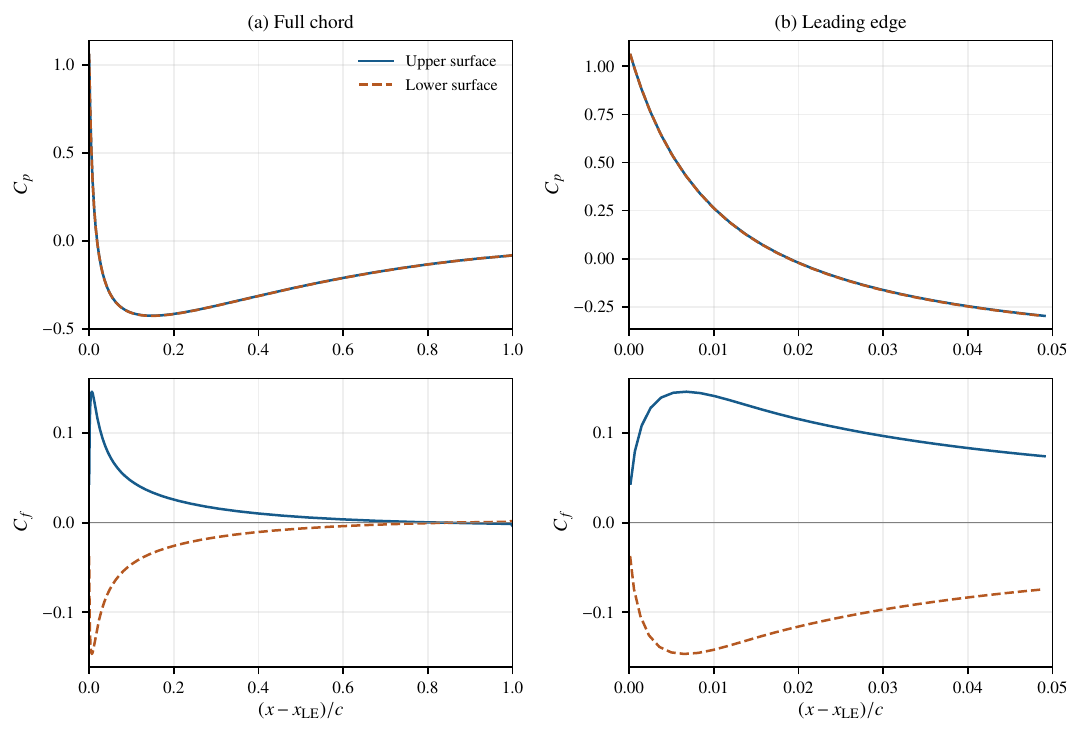}
\caption{NACA0012 NURBS solution: surface pressure coefficient $C_p$
(top) and signed skin-friction coefficient $C_f$ (bottom), over
(a)~the full chord and (b)~the leading-edge region.}
\label{fig:naca_surface}
\end{figure*}

Figure~\ref{fig:naca_mach} shows the boundary layer and downstream wake
through Mach-number contours. Figure~\ref{fig:naca_surface} presents the
surface pressure and skin-friction distributions, and
Table~\ref{tab:naca_drag} reports the integrated force and moment
coefficients.

\begin{table}[!tbp]
\centering
\caption{Integrated NACA0012 NURBS coefficients with
$L_{\mathrm{ref}}=c$ and moment center $(0,0)$.}
\label{tab:naca_drag}
\begingroup
\small
\setlength{\tabcolsep}{6pt}
\begin{tabular}{@{}rrrrr@{}}
\toprule
$C_{D,\mathrm p}$ & $C_{D,\mathrm v}$ & $C_D$ & $C_L$ & $C_M$\\
\midrule
0.02313329 & 0.03328674 & 0.05642004 & $-0.00059756$ & $-0.00024472$\\
\bottomrule
\end{tabular}
\endgroup
\end{table}

The surface response is concentrated near the leading edge, with
shear magnitude decreasing downstream. Viscous drag dominates, while
lift and moment remain small at zero incidence.

\subsection{Tandem NACA0012 Airfoils}
\label{sec:tandem_naca}

\begin{figure*}[tbp]
\centering
\includegraphics[width=0.98\textwidth]{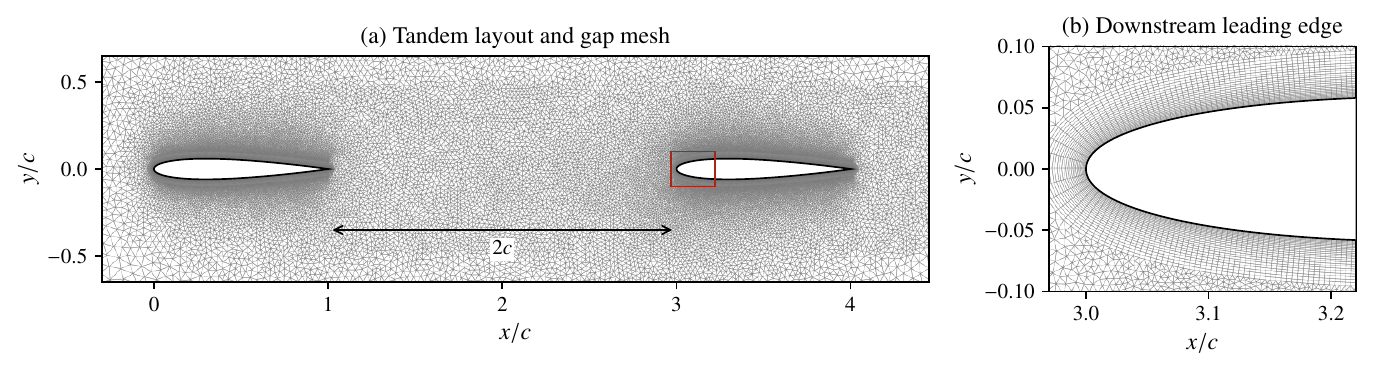}
\caption{Tandem NACA0012 on the finest mesh: (a)~airfoil arrangement and gap
mesh; (b)~downstream leading edge and stretched wall layer.}
\label{fig:tandem_mesh}
\end{figure*}

We consider two aligned NACA0012 airfoils with the profile and flow conditions of
Section~\ref{sec:naca}, with $M=0.5$ and $\mathrm{Re}=5000$ based on the
chord length. Both airfoils have no-slip adiabatic walls, with
inflow/outflow boundary conditions at the outer boundary.
Figure~\ref{fig:tandem_mesh} shows
the tandem arrangement and downstream leading-edge mesh.

Table~\ref{tab:tandem_mesh} compares lift and drag under joint
wall-tangential and normal refinement. The following flow comparisons
use the finest mesh.

\begin{table*}[tbp]
\centering
\caption{Tandem NACA0012 mesh comparison for cubic reconstruction. $N_w$ is the number of wall faces per airfoil; subscripts 1 and 2 denote the upstream and downstream bodies.}
\label{tab:tandem_mesh}
\begin{tabular}{@{}rrrrrr@{}}
\toprule
$N$ & $N_w$ & $C_{D,1}$ & $C_{D,2}$ & $C_{L,1}$ & $C_{L,2}$ \\
\midrule
13{,}178 & 160 & 0.05353617 & 0.04172633 & -0.00008150 & -0.00253194 \\
57{,}624 & 320 & 0.05462649 & 0.03804014 & 0.00001400 & 0.00031127 \\
123{,}404 & 480 & 0.05441015 & 0.03727878 & 0.00001516 & 0.00010482 \\
\bottomrule
\end{tabular}
\end{table*}

\begin{figure*}[tbp]
\centering
\includegraphics[width=0.98\textwidth]{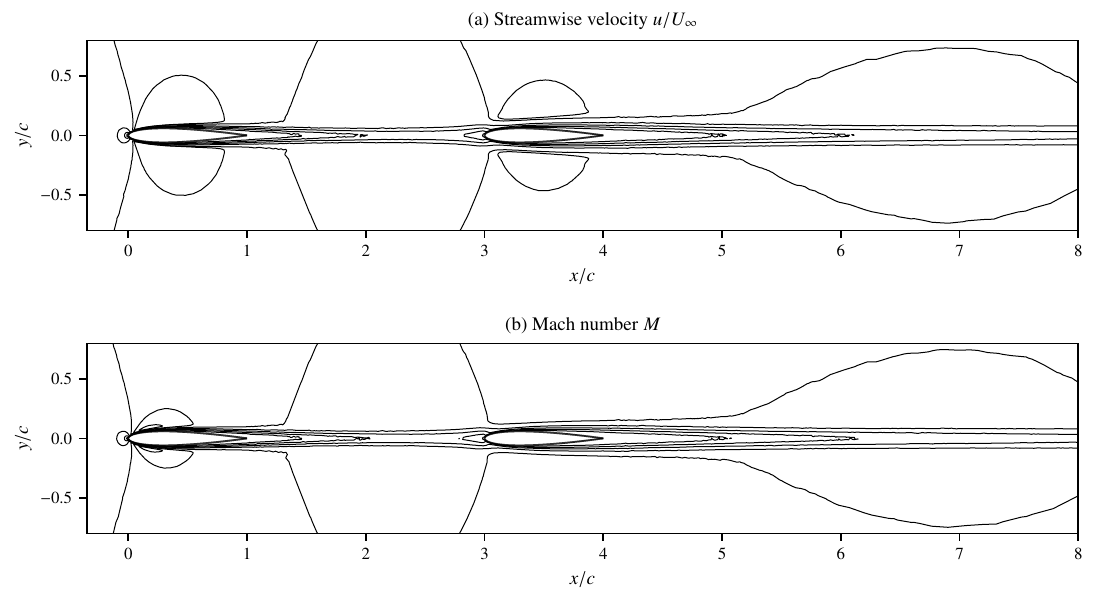}
\caption{Tandem NACA0012: contours of (a)~streamwise velocity $u/U_\infty$
and (b)~Mach number.}
\label{fig:tandem_flow}
\end{figure*}

\begin{figure*}[tbp]
\centering
\includegraphics[width=0.94\textwidth]{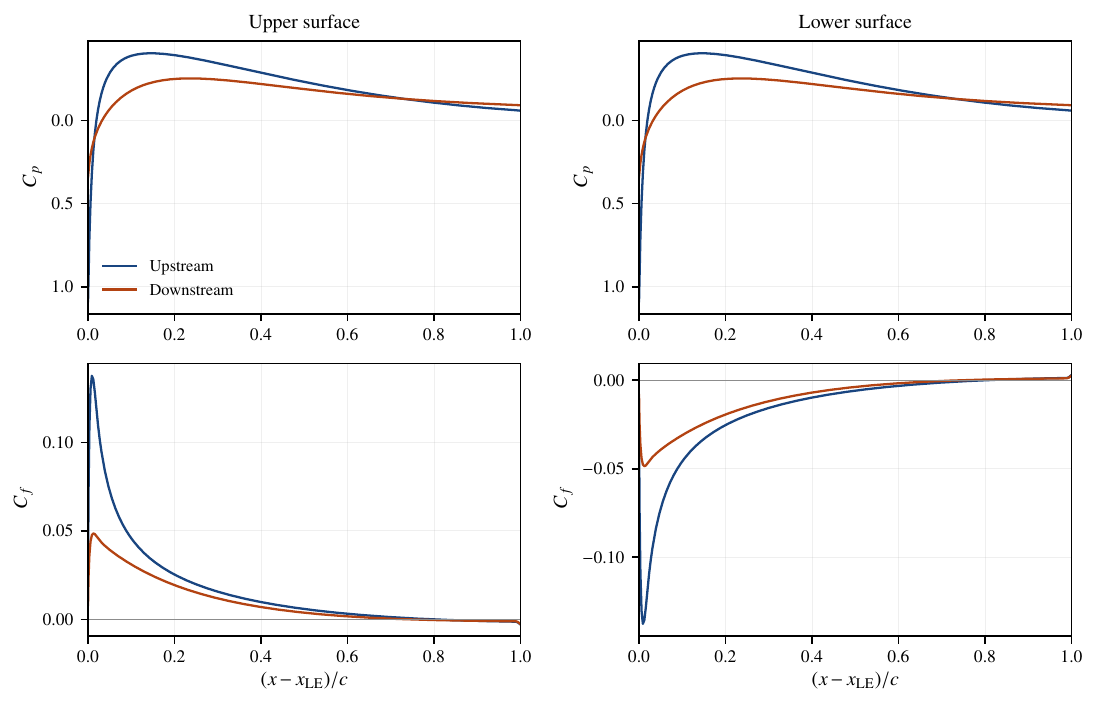}
\caption{Finest-mesh tandem NACA0012 solution:
pressure and signed skin-friction coefficients versus local chord
coordinate. Values are physical arc-length averages over each wall face.}
\label{fig:tandem_wall}
\end{figure*}

Figure~\ref{fig:tandem_flow} shows streamwise-velocity and Mach-number
contours. Figure~\ref{fig:tandem_wall} shows the surface pressure and
skin-friction distributions on the upstream and downstream airfoils.

Table~\ref{tab:tandem_forces} summarizes the integrated force and
moment coefficients.

\begin{table*}[tbp]
\centering
\caption{Tandem NACA0012 coefficients. The reference length is the unit airfoil chord; $C_{M,\mathrm{LE}}$ is the moment coefficient about the corresponding leading edge, positive counterclockwise.}
\label{tab:tandem_forces}
\begingroup
\small
\begin{tabular}{@{}lrrrrr@{}}
\toprule
Airfoil & $C_{D,\mathrm p}$ & $C_{D,\mathrm v}$ & $C_D$ & $C_L$ & $C_{M,\mathrm{LE}}$ \\
\midrule
Upstream & 0.02209983 & 0.03231032 & 0.05441015 & 0.00001516 & 0.00000509 \\
Downstream & 0.01802822 & 0.01925056 & 0.03727878 & 0.00010482 & 0.00003780 \\
\bottomrule
\end{tabular}
\endgroup
\end{table*}

The upstream wake reaches the downstream leading edge.
The downstream airfoil has lower drag than the upstream airfoil.

\subsection{Boundary-Layer Separation over a Gaussian Bump}
\label{sec:gaussian_bump}

\begin{figure*}[!tbp]
\centering
\includegraphics[width=0.8\textwidth]{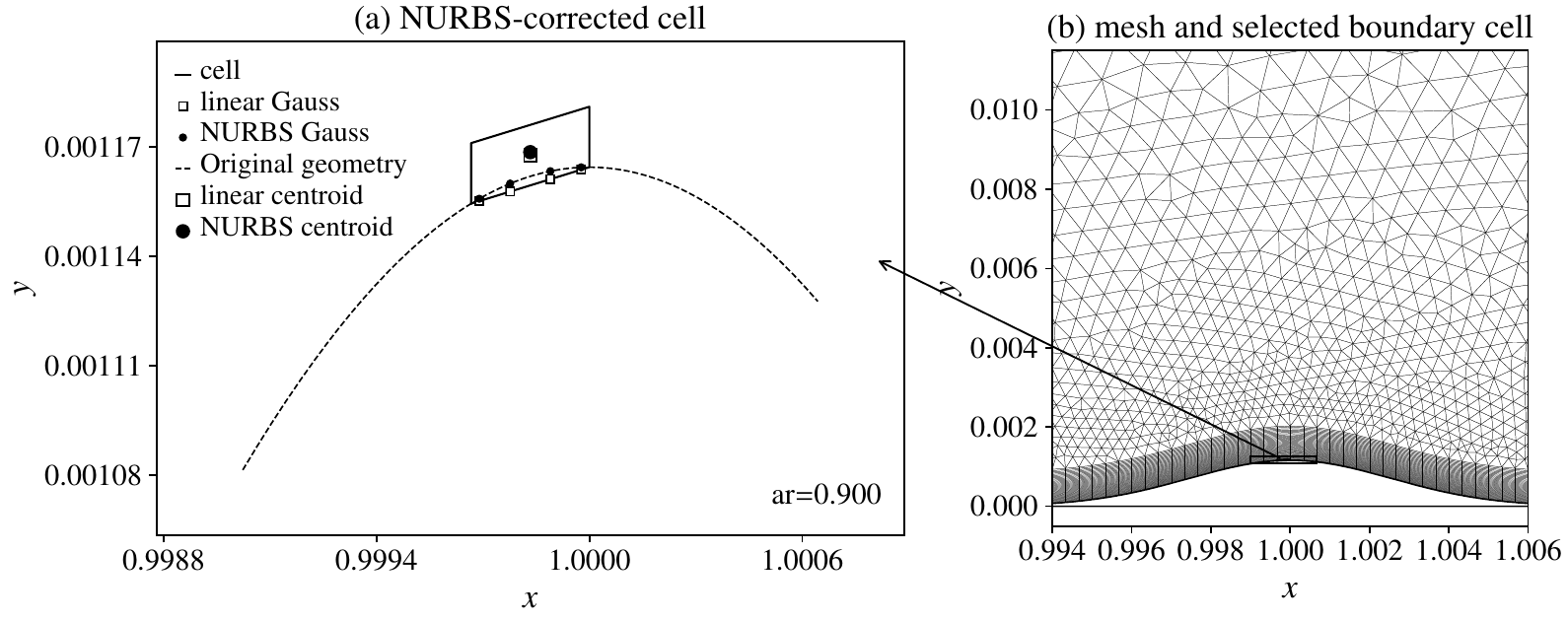}
\caption{Gaussian bump: (a)~polygonal and NURBS representations of a
near-crest boundary cell; (b)~local hybrid mesh. The area ratio compares
the curved cell with its straight-sided counterpart.}
\label{fig:gaussian_bump_mesh}
\end{figure*}

A Gaussian roughness element perturbs the flat-plate laminar boundary
layer at freestream Mach number $M=0.5$ and Reynolds number
$\mathrm{Re}=4\times10^5$. The roughness profile is
\begin{equation}
y_b(x) = h_r \exp\left[-\frac{(x-x_c)^2}{\beta^2}\right],
\label{eq:gaussian_bump}
\end{equation}
with $h_r=0.15\delta_{99}$, $\beta=0.0035922$, and $x_c=1$,
following Mengaldo \emph{et al.}~\cite{mengaldo2015}. Here $\delta_{99}$
is the Blasius boundary-layer thickness at the bump; the distance
from the leading edge to the bump defines the reference length and
$\mathrm{Re}$. At $x=0$, the lower boundary changes from a slip boundary to a
no-slip isothermal wall with $T_w=T_\infty$.
Inflow/outflow boundary conditions are applied at
the inlet, upper boundary, and outlet, using the freestream reference state.
Figure~\ref{fig:gaussian_bump_mesh} shows the stretched quadrilateral
wall layers and compares polygonal and NURBS representations of a
near-crest cell.

For comparison with the direct numerical simulation (DNS) and triple-deck results~\cite{mengaldo2015},
we use $\xi=Lx$, $L=0.05\ \mathrm m$, and the upstream reference station
$x_u=0.96$, with normalization
\begin{equation}
\begin{aligned}
\mathcal S&=\frac{C_f}{C_{f,u}}=\frac{\tau_{xy}}{\tau_{xy,u}},\\
\mathcal G&=\frac{1}{C_{f,u}}\frac{\mathrm dC_p}{\mathrm dx}
=\frac{L}{\tau_{xy,u}}\frac{\mathrm dP}{\mathrm d\xi}.
\end{aligned}
\label{eq:bump_reference_scaling}
\end{equation}
The subscript $u$ denotes the upstream shear for each method.

\begin{figure}[!tbp]
\centering
\includegraphics[width=0.95\textwidth]{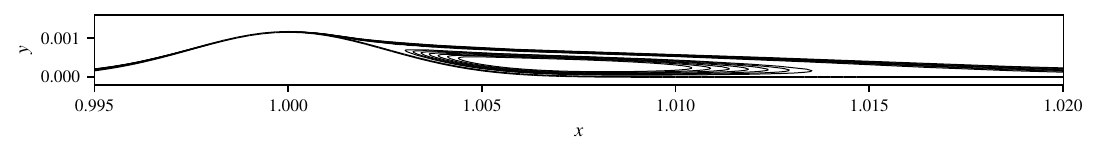}
\caption{Gaussian bump: Mach-number contours in the low-speed region downstream of the
crest. Both axes use the same length scale.}
\label{fig:gaussian_bump_mach}
\end{figure}

\begin{figure}[!tbp]
\centering
\includegraphics[width=0.80\textwidth]{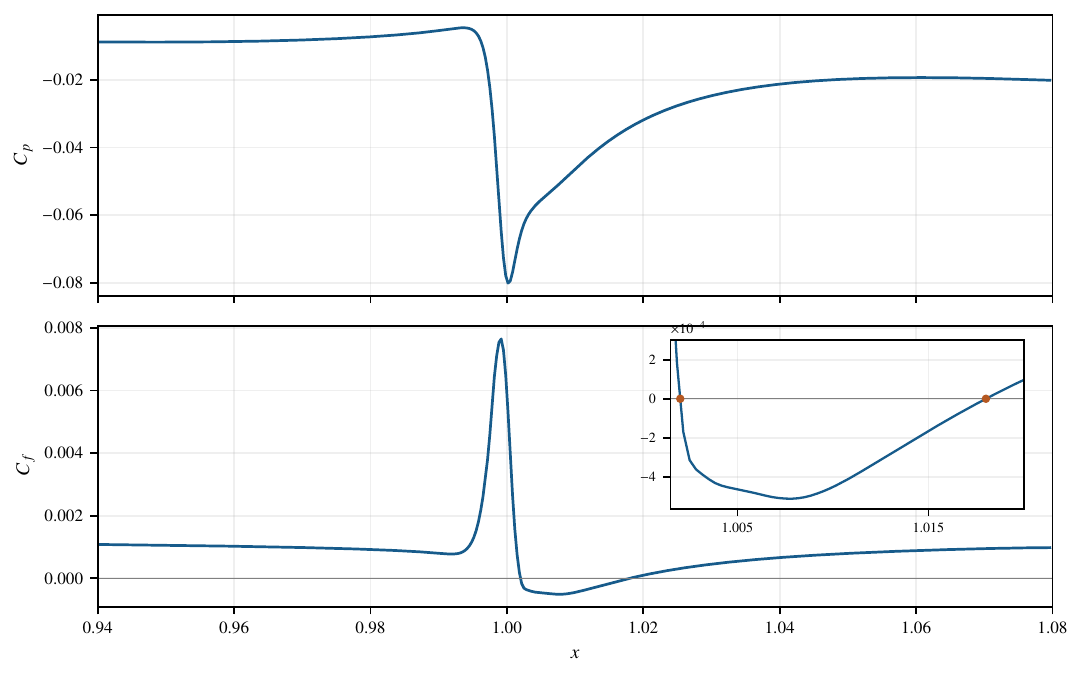}
\caption{Gaussian bump: $C_p$ (top) and signed $C_f$ (bottom).
The inset enlarges the negative-$C_f$ region; markers indicate separation
and reattachment.}
\label{fig:gaussian_bump_surface}
\end{figure}

Figure~\ref{fig:gaussian_bump_mach} shows the low-speed region downstream
of the crest. Figure~\ref{fig:gaussian_bump_surface} presents the local
pressure and skin-friction distributions; Table~\ref{tab:gaussian_bump_summary}
summarizes separation, reattachment, and drag.

\begin{figure*}[!tbp]
\centering
\includegraphics[width=0.80\textwidth]{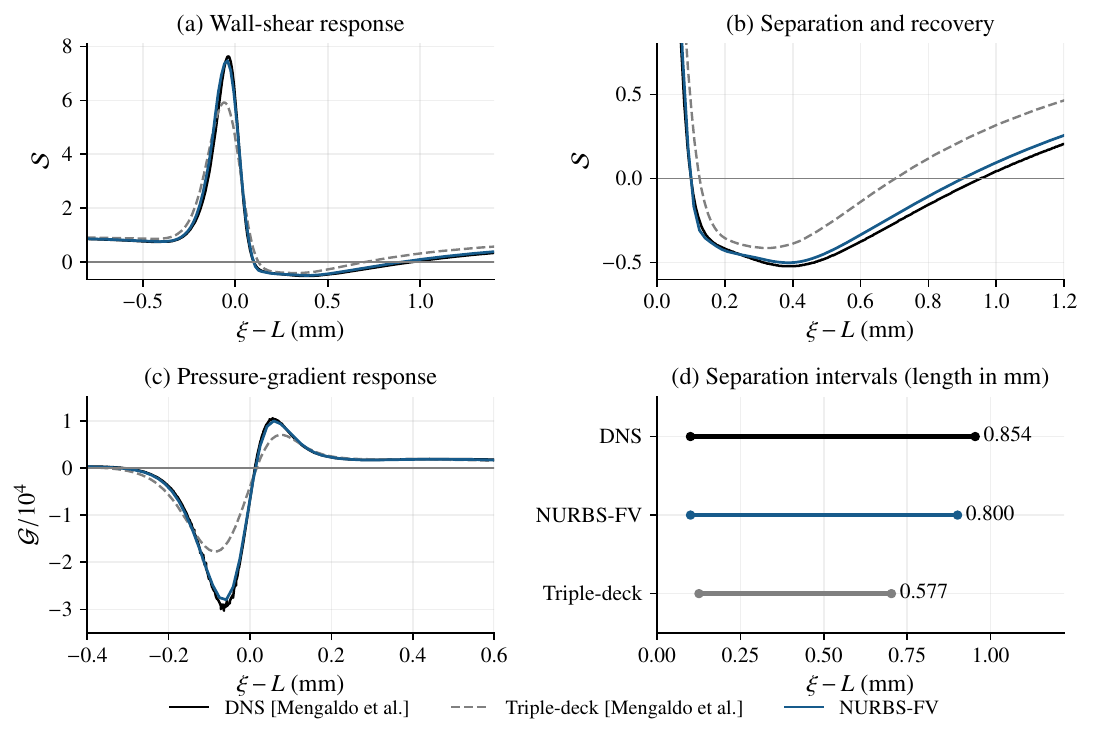}
\caption{Gaussian bump: comparison with the DNS and triple-deck results
of Mengaldo \emph{et al.}~\cite{mengaldo2015}: (a)~normalized wall shear,
(b)~separation and recovery, (c)~normalized pressure gradient, and
(d)~separation intervals. Normalization follows Eq.~\eqref{eq:bump_reference_scaling};
$\xi-L$ is the dimensional distance from the bump center.}
\label{fig:gaussian_bump_reference_comparison}
\end{figure*}

\begin{table*}[!tbp]
\centering
\caption{Gaussian bump: separation measures and drag coefficients on three hybrid meshes, with drag integrated over $0\le x\le1.25$ and $L_{\mathrm{ref}}=1$.}
\label{tab:gaussian_bump_summary}
\begingroup
\small
\setlength{\tabcolsep}{6pt}
\renewcommand{\arraystretch}{1.1}
\begin{tabular*}{0.9\textwidth}{@{\extracolsep{\fill}}rrrrrrr@{}}
\toprule
$N$ & $x_d$ & $x_r$ & $l_{\mathrm{sep}}$ & $C_{D,\mathrm p}$ & $C_{D,\mathrm v}$ & $C_D$\\
\midrule
25{,}391 & 1.00294 & 1.01243 & 0.00948 & $2.68599\times10^{-5}$ & $1.14190\times10^{-3}$ & $1.16876\times10^{-3}$\\
56{,}992 & 1.00219 & 1.01582 & 0.01363 & $3.43297\times10^{-5}$ & $1.44234\times10^{-3}$ & $1.47667\times10^{-3}$\\
128{,}226 & 1.00201 & 1.01802 & 0.01601 & $4.57586\times10^{-5}$ & $1.83918\times10^{-3}$ & $1.88494\times10^{-3}$\\
\bottomrule
\end{tabular*}
\endgroup
\end{table*}

Negative wall shear coincides with the low-speed region, identifying
the separation bubble behind the crest.
Figure~\ref{fig:gaussian_bump_reference_comparison} compares normalized
wall-shear and pressure-gradient responses with the DNS and triple-deck
curves. Table~\ref{tab:gaussian_bump_reference_comparison} gives the
corresponding separation measures and peak normalized shear.

\FloatBarrier
\begingroup
\setlength{\intextsep}{4pt}
\captionsetup{skip=4pt}
\begin{table}[H]
\centering
\caption{Gaussian bump: comparison of NURBS-FV with the published DNS and
triple-deck results~\cite{mengaldo2015}.
The coordinates $x_d$ and $x_r$ are nondimensional;
$L\,l_{\mathrm{sep}}$ is the physical separation length.}
\label{tab:gaussian_bump_reference_comparison}
\begingroup
\small
\setlength{\tabcolsep}{3pt}
\begin{tabular}{@{}lrrrr@{}}
\toprule
Method & $x_d$ & $x_r$ & $L\,l_{\mathrm{sep}}$ (mm) & $\max\mathcal{S}$\\
\midrule
NURBS-FV & 1.00201 & 1.01802 & 0.8004 & 7.474\\
DNS~\cite{mengaldo2015} & 1.00200 & 1.01907 & 0.8535 & 7.632\\
Triple-deck~\cite{mengaldo2015} & 1.00251 & 1.01405 & 0.5769 & 5.923\\
\bottomrule
\end{tabular}
\endgroup
\end{table}
\endgroup

\begin{samepage}
The fine-grid NURBS-FV predictions of separation length and peak normalized wall
shear are closer to the DNS values than the triple-deck predictions.
The predicted separation length is $6.2\%$ shorter than the DNS value.
\par
\end{samepage}

\FloatBarrier
\section{Conclusions}

This work has developed a NURBS-enhanced high-order $k$-exact finite-volume
method for steady compressible laminar boundary layers on curved domains with
triangular--quadrilateral hybrid meshes. The proposed wall boundary treatment
employs sequential reconstruction to couple physical wall conditions prescribed in primitive variables with the
$k$-exact reconstruction of conservative variables, removing the need to match
the reconstruction variables to those used to specify the boundary conditions.
Its combination with fitted NURBS geometry, curved-cell reconstruction moments,
and the numerical viscous fluxes establishes a consistent spatial framework
linking curved-domain representation, near-wall approximation, and the
finite-volume flux balance.

Within this framework, geometric fidelity enters both surface integration and
curved-cell reconstruction moments, while physical wall
conditions are incorporated through boundary reconstruction goals
and numerical wall fluxes. The resulting discretization connects curved
geometry, high-order reconstruction, viscous coupling, and wall treatment
within a unified finite-volume formulation. The demonstrated
resolution of compressible laminar boundary layers and their wall response shows
that this formulation can support the evaluation of near-wall flow structures,
surface stresses, aerodynamic loads, and separation behavior without reducing
these quantities to isolated boundary corrections.

These capabilities provide a numerical foundation for high-fidelity viscous-flow
simulation on curved domains. Beyond the configurations considered here, the
formulation offers a basis for extending high-order finite-volume analysis to
more complex aerodynamic geometries while retaining the connection between
resolved near-wall structures and surface response. Such a capability is
relevant to predictive aerodynamic analysis and design, where geometry-dependent
viscous effects must be evaluated reliably as part of a larger computational
workflow. The present work therefore represents a step toward high-fidelity
compressible-flow simulation and the future use of high-order flow solvers as
analysis components in multidisciplinary design analysis and optimization
(MDAO).

\section*{CRediT authorship contribution statement}
\textbf{Yisu Li:} Writing -- original draft, Visualization, Software,
Methodology, Formal analysis.
\textbf{Xucheng Meng:} Writing -- review \& editing, Supervision,
Conceptualization.
\textbf{Guanghui Hu:} Writing -- review \& editing, Supervision,
Project administration, Conceptualization.

\section*{Declaration of generative AI and AI-assisted technologies in the manuscript preparation process}
During the preparation of this work, Yisu Li used Codex (based on GPT-5)
for coding assistance, syntax checking, and language polishing. After using
this tool, the authors reviewed and edited
the content as needed and take full responsibility for the content of the
published article.

\section*{Declaration of competing interest}
The authors declare that they have no known competing financial interests
or personal relationships that could have appeared to influence the work
reported in this paper.

\nocite{*}
\begingroup
\footnotesize
\bibliographystyle{unsrt}
\bibliography{src/references}

@techreport{slotnick2014,
  author = {J. P. Slotnick and A. Khodadoust and J. Alonso and D. Darmofal and W. Gropp and E. Lurie and D. J. Mavriplis},
  title = {{CFD Vision 2030 Study}: A Path to Revolutionary Computational Aerosciences},
  institution = {{NASA}},
  number = {NASA/CR-2014-218178},
  year = {2014},
}

@article{Sevilla2008Euler,
author = {Sevilla, R. and Fernández-Méndez, S. and Huerta, A.},
title = {{NURBS-enhanced finite element method for Euler equations}},
journal = {International Journal for Numerical Methods in Fluids},
volume = {57},
number = {9},
pages = {1051-1069},
year = {2008}
}

@article{rai1993,
  author = {M. M. Rai and P. Moin},
  title = {Direct numerical simulation of transition and turbulence in a spatially evolving boundary layer},
  journal = {J. Comput. Phys.},
  volume = {109},
  number = {2},
  pages = {169--192},
  year = {1993},
}

@article{alam2000,
  author = {M. Alam and N. D. Sandham},
  title = {Direct numerical simulation of `short' laminar separation bubbles with turbulent reattachment},
  journal = {J. Fluid Mech.},
  volume = {410},
  pages = {1--28},
  year = {2000},
}

@article{jones2008,
  author = {L. E. Jones and R. D. Sandberg and N. D. Sandham},
  title = {Direct numerical simulations of forced and unforced separation bubbles on an airfoil at incidence},
  journal = {J. Fluid Mech.},
  volume = {602},
  pages = {175--207},
  year = {2008},
}

@article{mengaldo2015,
  author = {G. Mengaldo and M. Kravtsova and A. I. Ruban and S. J. Sherwin},
  title = {Triple-deck and direct numerical simulation analyses of high-speed subsonic flows past a roughness element},
  journal = {J. Fluid Mech.},
  volume = {774},
  pages = {311--323},
  year = {2015},
}

@article{wang2013high,
  author = {Z. J. Wang and K. Fidkowski and R. Abgrall and F. Bassi and D. Caraeni and A. Cary and H. Deconinck and R. Hartmann and K. Hillewaert and H. T. Huynh and others},
  title = {High-order {CFD} methods: current status and perspective},
  journal = {Int. J. Numer. Methods Fluids},
  volume = {72},
  number = {8},
  pages = {811--845},
  year = {2013},
}

@article{wan2025curved,
  author = {Y. Wan and S. Liu and Y. Xia},
  title = {Convergence of the hybrid {WENO} scheme for steady compressible {Navier--Stokes} equations in curved geometries using {Cartesian} grids},
  journal = {J. Sci. Comput.},
  volume = {103},
  number = {2},
  pages = {61},
  year = {2025},
}

@article{navah2020verification,
  author = {F. Navah and S. Nadarajah},
  title = {On the verification of {CFD} solvers of all orders of accuracy on curved wall-bounded domains and for realistic {RANS} flows},
  journal = {Comput. Fluids},
  volume = {205},
  pages = {104504},
  year = {2020},
  doi = {10.1016/j.compfluid.2020.104504},
}

@article{bassi1997,
  author = {F. Bassi and S. Rebay},
  title = {A high-order accurate discontinuous finite element method for the numerical solution of the compressible {Navier--Stokes} equations},
  journal = {J. Comput. Phys.},
  volume = {131},
  number = {2},
  pages = {267--279},
  year = {1997},
}

@article{cockburn2001,
  author = {B. Cockburn and C.-W. Shu},
  title = {Runge--{Kutta} discontinuous {Galerkin} methods for convection-dominated problems},
  journal = {J. Sci. Comput.},
  volume = {16},
  number = {3},
  pages = {173--261},
  year = {2001},
}

@article{yan2021implicit,
  author = {Z.-G. Yan and Y. Pan and G. Castiglioni and K. Hillewaert and J. Peir\'{o} and D. Moxey and S. J. Sherwin},
  title = {{Nektar++}: Design and implementation of an implicit, spectral/$hp$ element, compressible flow solver using a {Jacobian}-free {Newton} {Krylov} approach},
  journal = {Comput. Math. Appl.},
  volume = {81},
  pages = {351--372},
  year = {2021},
}

@techreport{huynh2007,
  author = {H. T. Huynh},
  title = {A flux reconstruction approach to high-order schemes including discontinuous {Galerkin} methods},
  institution = {{AIAA}},
  type = {Paper},
  number = {2007-4079},
  year = {2007},
}

@article{wang2009,
  author = {Z. J. Wang and H. Gao},
  title = {A unifying lifting collocation penalty formulation including the discontinuous {Galerkin}, spectral volume/difference methods for conservation laws on mixed grids},
  journal = {J. Comput. Phys.},
  volume = {228},
  number = {21},
  pages = {8161--8186},
  year = {2009},
}

@article{dhib2024implicit,
  author = {R. Dhib and F. {Ben Ameur} and R. Vandenhoeck and A. Lani and S. Poedts},
  title = {Development of an implicit high-order {Flux Reconstruction} solver for high-speed flows on simplex elements},
  journal = {Comput. Phys. Commun.},
  volume = {295},
  pages = {109006},
  year = {2024},
}

@article{wang2007sd,
  author = {Z. J. Wang and Y. Liu and G. May and A. Jameson},
  title = {Spectral difference method for unstructured grids {II}: Extension to the {Euler} equations},
  journal = {J. Sci. Comput.},
  volume = {32},
  number = {1},
  pages = {45--71},
  year = {2007},
}

@techreport{barth1990,
  author = {T. J. Barth and P. O. Frederickson},
  title = {Higher order solution of the {Euler} equations on unstructured grids using quadratic reconstruction},
  institution = {{AIAA}},
  type = {Paper},
  number = {90-0013},
  year = {1990},
}

@article{olliviergooch2002,
  author = {C. Ollivier-Gooch and Van Altena, M.},
  title = {A high-order-accurate unstructured mesh finite-volume scheme for the advection--diffusion equation},
  journal = {J. Comput. Phys.},
  volume = {181},
  number = {2},
  pages = {729--752},
  year = {2002},
}

@inproceedings{michalak2009,
  author = {C. Michalak and C. Ollivier-Gooch},
  title = {Unstructured high-order accurate finite-volume solutions of the {Navier--Stokes} equations},
  booktitle = {47th AIAA Aerospace Sciences Meeting},
  address = {Orlando, FL},
  note = {{AIAA} Paper 2009-954},
  year = {2009},
}

@article{hughes2005,
  author = {T. J. R. Hughes and J. A. Cottrell and Y. Bazilevs},
  title = {Isogeometric analysis: {CAD}, finite elements, {NURBS}, exact geometry and mesh refinement},
  journal = {Comput. Methods Appl. Mech. Engrg.},
  volume = {194},
  number = {39--41},
  pages = {4135--4195},
  year = {2005},
}

@article{sevilla2008,
  author = {R. Sevilla and S. Fern\'{a}ndez-M\'{e}ndez and A. Huerta},
  title = {{NURBS}-enhanced finite element method ({NEFEM})},
  journal = {Int. J. Numer. Methods Eng.},
  volume = {76},
  number = {1},
  pages = {56--83},
  year = {2008},
}

@article{menghu2018,
  author = {X. Meng and G. Hu},
  title = {A {NURBS}-enhanced finite volume solver for steady {Euler} equations},
  journal = {J. Comput. Phys.},
  volume = {359},
  pages = {77--92},
  year = {2018},
}

@article{menghu2022adaptive,
  author = {X. Meng and G. Hu},
  title = {A {NURBS}-enhanced finite volume method for steady {Euler} equations with goal-oriented $h$-adaptivity},
  journal = {Commun. Comput. Phys.},
  volume = {32},
  number = {2},
  pages = {490--523},
  year = {2022},
}

@article{costa2019,
  author = {R. Costa and J. M. N\'obrega and S. Clain and G. J. Machado and R. Loub\`ere},
  title = {Very high-order accurate finite volume scheme for the convection--diffusion equation with general boundary conditions on arbitrary curved boundaries},
  journal = {Int. J. Numer. Methods Eng.},
  volume = {117},
  number = {2},
  pages = {188--220},
  year = {2019},
}

@article{jalali2017rans,
  author = {A. Jalali and C. Ollivier-Gooch},
  title = {Higher-order unstructured finite volume {RANS} solution of turbulent compressible flows},
  journal = {Comput. Fluids},
  volume = {143},
  pages = {32--47},
  year = {2017},
}

@article{huang2022mixed,
  author = {Q.-M. Huang and Y.-X. Ren and Q. Wang and J.-H. Pan},
  title = {High-order compact finite volume schemes for solving the {Reynolds} averaged {Navier--Stokes} equations on the unstructured mixed grids with a large aspect ratio},
  journal = {J. Comput. Phys.},
  volume = {467},
  pages = {111458},
  year = {2022},
}

@article{gassner2007,
  author = {G. Gassner and F. L\"orcher and C.-D. Munz},
  title = {A contribution to the construction of diffusion fluxes for finite volume and discontinuous {Galerkin} schemes},
  journal = {J. Comput. Phys.},
  volume = {224},
  number = {2},
  pages = {1049--1063},
  year = {2007},
}

@article{wang2017variational,
  author = {Q. Wang and Y.-X. Ren and J. Pan and W. Li},
  title = {Compact high order finite volume method on unstructured grids {III}: Variational reconstruction},
  journal = {J. Comput. Phys.},
  volume = {337},
  pages = {1--26},
  year = {2017},
}

@article{thompson1987,
  author = {K. W. Thompson},
  title = {Time dependent boundary conditions for hyperbolic systems},
  journal = {J. Comput. Phys.},
  volume = {68},
  number = {1},
  pages = {1--24},
  year = {1987},
}

@article{giles1990,
  author = {M. B. Giles},
  title = {Nonreflecting boundary conditions for {Euler} equation calculations},
  journal = {AIAA J.},
  volume = {28},
  number = {12},
  pages = {2050--2058},
  year = {1990},
}

@article{poinsot1992,
  author = {T. J. Poinsot and S. K. Lele},
  title = {Boundary conditions for direct simulations of compressible viscous flows},
  journal = {J. Comput. Phys.},
  volume = {101},
  number = {1},
  pages = {104--129},
  year = {1992},
}

@inproceedings{huang2018wellposed,
  author = {A. C. Huang and S. R. Allmaras and M. C. Galbraith and D. L. Darmofal},
  title = {Well-posed subsonic inflow/outflow boundary conditions for the {Navier--Stokes} equations},
  booktitle = {2018 AIAA Aerospace Sciences Meeting},
  note = {{AIAA} Paper 2018-0361},
  year = {2018},
  doi = {10.2514/6.2018-0361},
}

@article{nordstrom1995,
  author = {J. Nordstr\"om},
  title = {The use of characteristic boundary conditions for the {Navier--Stokes} equations},
  journal = {Comput. Fluids},
  volume = {24},
  number = {5},
  pages = {609--623},
  year = {1995},
}

@article{lodato2008,
  author = {G. Lodato and P. Domingo and L. Vervisch},
  title = {Three-dimensional boundary conditions for direct and large-eddy simulation of compressible viscous flows},
  journal = {J. Comput. Phys.},
  volume = {227},
  number = {10},
  pages = {5105--5143},
  year = {2008},
}

@article{motheau2017,
  author = {E. Motheau and A. Almgren and J. B. Bell},
  title = {{Navier--Stokes} characteristic boundary conditions using ghost cells},
  journal = {AIAA J.},
  volume = {55},
  number = {10},
  pages = {3399--3408},
  year = {2017},
}

@article{chassaing2013,
  author = {J.-C. Chassaing and X. Nogueira and S. Khelladi},
  title = {Moving {Kriging} reconstruction for high-order finite volume computation of compressible flows},
  journal = {Comput. Methods Appl. Mech. Engrg.},
  volume = {253},
  pages = {463--478},
  year = {2013},
}

@article{menasria2019,
  author = {A. Menasria and P. Brenner and P. Cinnella},
  title = {Improving the treatment of near-wall regions for multiple-correction $k$-exact schemes},
  journal = {Comput. Fluids},
  volume = {181},
  pages = {116--134},
  year = {2019},
}

@article{mccorquodale2011,
  author = {P. McCorquodale and P. Colella},
  title = {A high-order finite-volume method for conservation laws on locally refined grids},
  journal = {Commun. Appl. Math. Comput. Sci.},
  volume = {6},
  number = {1},
  pages = {1--25},
  year = {2011},
}

@article{zanotti2016,
  author = {O. Zanotti and M. Dumbser},
  title = {Efficient conservative {ADER} schemes based on {WENO} reconstruction and space-time predictor in primitive variables},
  journal = {Comput. Astrophys. Cosmol.},
  volume = {3},
  pages = {1},
  year = {2016},
}

@article{roe1981,
  author = {P. L. Roe},
  title = {Approximate {Riemann} solvers, parameter vectors, and difference schemes},
  journal = {J. Comput. Phys.},
  volume = {43},
  number = {2},
  pages = {357--372},
  year = {1981},
}

@article{toro1994,
  author = {E. F. Toro and M. Spruce and W. Speares},
  title = {Restoration of the contact surface in the {HLL}-{Riemann} solver},
  journal = {Shock Waves},
  volume = {4},
  number = {1},
  pages = {25--34},
  year = {1994},
}

@article{venditti2003,
  author = {D. A. Venditti and D. L. Darmofal},
  title = {Anisotropic grid adaptation for functional outputs: application to two-dimensional viscous flows},
  journal = {J. Comput. Phys.},
  volume = {187},
  number = {1},
  pages = {22--46},
  year = {2003},
}

@article{moro2017,
  author = {D. Moro and N. C. Nguyen and J. Peraire and M. Drela},
  title = {Mesh topology preserving boundary-layer adaptivity method for steady viscous flows},
  journal = {AIAA J.},
  volume = {55},
  number = {6},
  pages = {1970--1985},
  year = {2017},
}

@article{dussin2009,
  author = {D. Dussin and M. Fossati and A. Guardone and L. Vigevano},
  title = {Hybrid grid generation for two-dimensional high-{Reynolds} flows},
  journal = {Comput. Fluids},
  volume = {38},
  number = {10},
  pages = {1863--1875},
  year = {2009},
}

@article{antoniadis2017,
  author = {A. F. Antoniadis and P. Tsoutsanis and D. Drikakis},
  title = {Assessment of high-order finite volume methods on unstructured meshes for {RANS} solutions of aeronautical configurations},
  journal = {Comput. Fluids},
  volume = {146},
  pages = {86--104},
  year = {2017},
}

@book{piegl1997,
  author = {L. Piegl and W. Tiller},
  title = {The {NURBS} Book},
  edition = {second},
  publisher = {Springer-Verlag},
  address = {New York},
  year = {1997},
}

@article{koren1990,
  author = {B. Koren},
  title = {Multigrid and defect correction for the steady {Navier--Stokes} equations},
  journal = {J. Comput. Phys.},
  volume = {87},
  number = {1},
  pages = {25--46},
  year = {1990},
}

@article{kelley1998,
  author = {C. T. Kelley and D. E. Keyes},
  title = {Convergence analysis of pseudo-transient continuation},
  journal = {SIAM J. Numer. Anal.},
  volume = {35},
  number = {2},
  pages = {508--523},
  year = {1998},
}

@article{nejat2008,
  author = {A. Nejat and C. Ollivier-Gooch},
  title = {A high-order accurate unstructured finite volume {Newton}--{Krylov} algorithm for inviscid compressible flows},
  journal = {J. Comput. Phys.},
  volume = {227},
  number = {4},
  pages = {2582--2609},
  year = {2008},
}

@article{nejat2008preconditioning,
  author = {A. Nejat and C. Ollivier-Gooch},
  title = {Effect of discretization order on preconditioning and convergence of a high-order unstructured {Newton}--{GMRES} solver for the {Euler} equations},
  journal = {J. Comput. Phys.},
  volume = {227},
  number = {4},
  pages = {2366--2386},
  year = {2008},
}

@article{knoll2004,
  author = {D. A. Knoll and D. E. Keyes},
  title = {Jacobian-free {Newton}--{Krylov} methods: a survey of approaches and applications},
  journal = {J. Comput. Phys.},
  volume = {193},
  number = {2},
  pages = {357--397},
  year = {2004},
}

@mastersthesis{hoshyari2017,
  author = {S. Hoshyari},
  title = {A higher-order unstructured finite volume solver for three-dimensional compressible flows},
  school = {University of British Columbia},
  year = {2017},
}

@inproceedings{menasria2018,
  author = {A. Menasria and P. Brenner and P. Cinnella and G. Pont},
  title = {Toward an improved wall treatment for multiple-correction $k$-exact schemes},
  booktitle = {2018 Fluid Dynamics Conference},
  address = {Atlanta, GA},
  note = {{AIAA} Paper 2018-4164},
  year = {2018},
}

@article{jalali2014diffusive,
  author = {A. Jalali and M. Sharbatdar and C. Ollivier-Gooch},
  title = {Accuracy analysis of unstructured finite volume discretization schemes for diffusive fluxes},
  journal = {Comput. Fluids},
  volume = {101},
  pages = {220--232},
  year = {2014},
}

@article{sejekan2016diffusive,
  author = {C. Sejekan and C. Ollivier-Gooch},
  title = {Improving finite-volume diffusive fluxes through better reconstruction},
  journal = {Comput. Fluids},
  volume = {139},
  pages = {216--232},
  year = {2016},
}

@article{setzwein2024viscous,
  author = {F. Setzwein and P. Ess and P. Gerlinger},
  title = {Investigating a high-order viscous flux scheme for unstructured grids},
  journal = {AIAA J.},
  volume = {62},
  number = {12},
  pages = {4798--4811},
  year = {2024},
}
\endgroup

\end{document}